\documentclass[12pt,preprint]{aastex}

\shorttitle{RESOLVED WHITE DWARF -- RED DWARF SYSTEMS}
\shortauthors{Farihi, Wachter, \& Hoard}

\begin{document}

\title{WHITE DWARF -- RED DWARF SYSTEMS
	RESOLVED WITH THE {\em HUBBLE SPACE
	TELESCOPE}: I. FIRST RESULTS}

\author{J. Farihi\altaffilmark{1,2},
	 D. W. Hoard\altaffilmark{3,4}, \& 
	 S. Wachter\altaffilmark{3}}

\altaffiltext{1}{Gemini Observatory,
			Northern Operations,
			670 North A'ohoku Place,
			Hilo, HI 96720}
\altaffiltext{2}{Department of Physics \& Astronomy,
			University of California,
			430 Portola Plaza,
			Los Angeles, CA 90095}
\altaffiltext{3}{Spitzer Science Center,
			California Institute of Technology,
			MS 220-6,
			Pasadena, CA 91125}
\altaffiltext{4}{Department of Physics \& Astronomy,
			Pomona College,
			610 North College Ave,
			Claremont, CA 91711}


\email{jfarihi@gemini.edu,hoard@ipac.caltech.edu,wachter@ipac.caltech.edu}

\begin{abstract}

First results are presented for a {\em Hubble Space
Telescope} Advanced Camera for Surveys snapshot study
of white dwarfs with likely red dwarf companions.  Of
48 targets observed and analyzed so far, 27 are totally
or partially resolved into two or more components, while
an additional 15 systems are almost certainly unresolved
binaries.  These results provide the first direct empirical
evidence for a bimodal distribution of orbital separations
among binary systems containing at least one white dwarf.

\end{abstract}


\section{INTRODUCTION}

The study of low mass stellar and substellar companions
to white dwarfs yields useful information regarding the
initial mass function near the bottom of the main sequence
and below, the overall binary fraction of intermediate
mass stars, and the long term stability and survivability
of low mass objects in orbit about post-asymptotic giant
branch stars \citep{zuc87,zuc92,sch96,gre00,wac03,far04,
far05}.  Of particular interest is common envelope evolution
and its consequences for any low mass companion.  Understanding
how low mass, unevolved companions fare inside and outside a
common envelope is an easier task than distentangling binaries
which have experienced two envelopes, such as double degenerates,
and should provide insight into more complex binary evolution.

This paper presents the first results from a {\em HST}
Advanced Camera for Surveys (ACS) imaging survey of 90
candidate white dwarf + red dwarf binaries.  The goals
of the study are to empirically test the bimodal distribution
of orbital semimajor axes predicted by post-asymptotic giant
branch binary evolution models, and to examine the distribution
of companion masses as a function of current separation 
\citep{jea24,bon85,zuc87,val88,bon90,dek93,yun93,sch96,
liv96,max98,sch03,far04}.  

\section{PROGRAM DESCRIPTION}

\subsection{Primary Motivation}

The prime focus of the present study is to image white
dwarf + red dwarf pairs with sufficient spatial resolution
to directly probe the $\sim0.1-10$ AU range.  Specifically,
models of orbital expanion due to adiabatic mass loss, combined
with models of frictional inspiral within a circumbinary envelope, 
predict a gap between diminished ($a\la0.1$ AU, $P\sim10^{-2}-10^{-4}$
yr) and augmented ($a\ga5$ AU, $P\sim10^{1}-10^{3}$ yr) orbits for low
mass, unevolved companions to white dwarfs.  A crude model of the 
expected distribution is given in \citet{far04}; basically, companions
originally within $r=2$ AU are brought inward of 0.1 AU, while those 
formerly outside this radius migrate farther by a factor of 3.

\subsection{Sample Stars}

The selected program stars, taken from and described
in \citet{wac03}, are almost exclusively white dwarfs in
\citet{mcc99} which were found by \citet{wac03} to exhibit
excess near-infrared emission in the 2MASS point source
catalog \citep{cut03}.  Several additional targets of a
similar nature were taken from \citet{far04}.  White dwarfs
with measured near-infrared flux above the photospheric
contribution are strong candidates for harboring cool, low
mass companions.  With radii of $R\sim1$ $R_{Jupiter}$, low
mass stellar and substellar companions to white  dwarfs
($R\sim1$ $R_{\oplus}$) can easily dominate the spectral
energy distribution of the binary at red to near-infrared
wavelengths, despite their low luminosities and low effective
temperatures \citep{pro83,zuc87,zuc87b,zuc92,gre00,wac03,
far04,far05}.  The target list for the white dwarfs discussed
in this paper is given in Table \ref{tbl1}.

\subsection{Observations}

The present data were taken with the ACS \citep{for98} High
Resolution Channel (HRC) aboard {\em HST}.  The observations
at each target consisted of a small 4-point dither pattern with
the F814W (approximately $I$ band) filter to such a depth as to
reach signal-to-noise (S/N) $\ga50$ on both the white dwarf and its
suspected red dwarf companion.  These dithered frames were processed
with MULTIDRIZZLE\footnote{http://stsdas.stsci.edu/multidrizzle/}
to create a single combined, cosmic ray free image on which to
perform astrometry and photometry.

\subsection{Data Analysis}

Each reduced image was visually inspected, and if two or more
components were seen separated by $\ga0.''5$, or if the candidate
binary appeared to be a single unresolved point source, the analysis
was as follows.  For these single or well resolved multiple point
sources, astrometry and aperture photometry were performed with
standard IRAF tasks.  A few astrometric tasks were carried out
with IDP3\footnote{http://nicmos.as.arizona.edu/software/}, which 
has the ability to determine the length and orientation of the major
and minor axes in a gaussian profile fit.  Both centroid and radial
profile fits were executed for each star to determine its coordinates
on the chip and to measure the full width at half maximum (FWHM), as
well as eccentricity.  Radial profiles were fit using a $0.''075$ (3
pixel) radius, while photometry was carried out with a $0.''125$ (5
pixel) aperture radius.  Corrections to the standard ACS aperture, as
well as appropriate sky annuli values, and color corrections, were
taken from \citet{sir05}.  Counts were converted into flux, then 
into Vega magnitudes, and finally into Cousins $I$ band magnitudes,
all following the methods and calibrations of \citet{sir05}.

If a binary appeared resolved but with point-spread functions
(PSFs) separated by $\la0.''5$, then a different approach was
used.  In order to perform astrometry and photometry on these
components with overlapping PSFs, it was first necessary to remove
the nearby companion flux individually for each star.  For this
purpose, TinyTim\footnote{http://www.stsci.edu/software/tinytim/}
was used to generate ACS PSFs which were then processed with
MULTIDRIZZLE.  These multidrizzled PSFs were interpolated, 
scaled, and subtracted in turn from each star of the binary
systems to isolate one star at a time.  Each binary component
was then examined in the same manner as for single sources and
resolved components as described above.  However, after companion 
removal there were often residuals still contaminating aperture
photometry out to $r=5$ pixels.  For these close binary components,
an aperture of $r=2-3$ pixels was used and the measured flux 
was corrected to $r=5$ pixels using  aperture corrections
derived from all well-separated single point source science 
targets in the data set.  

\subsection{PSF Subtraction}

PSF subtraction was performed in the manner described
above for all science target point sources to search
for faint, close companions.  In a typical PSF subtracted
image, the residuals contain some small structure which is
likely attributable to imperfections in the artificial,
multidrizzled PSFs.  Although the residuals were typically
$<2-3$\% of the peak flux, this still might obscure any 
close companions at $\Delta m\ga4$ magnitudes.  There is
an ongoing effort to improve the PSF subtractions by: 1)
performing the subtractions on the calibrated, flat fielded
images and then multidrizzling the resultant images or; 2)
using real ACS multidrizzled PSFs rather than artificial
ones.

\section{FIRST RESULTS}

Table \ref{tbl1} lists all astrometry and photometry
results for the 48 targets analyzed so far in the program.
The table is organized so that the first line for a given
target provides data on the white dwarf and subsequent lines
are the companion data.  The first column lists the white dwarf
number from \citet{mcc99}, and the second column lists an alternate
name for the white dwarf or the companion name.  The third column
contains ``Yes'' if the components are totally or partially resolved
in the ACS observation, or ``No'' if unresolved.  The fourth column
lists the FWHM in arcseconds for the Airy disk from gaussian profile
fitting, and the fifth column lists the ratio of the major to minor
axes of the profile fit.  The sixth and seventh columns contain the
separation in arcseconds and the position angle in degrees of the
companion, if present.  The eighth and ninth columns list the
photometry, in Vega magnitudes, for the F814W filter and
Cousins $I$ band.  The last column lists system specific
notes.

\subsection{Photometric \& Astrometric Errors}

There are a several sources of photometric and astrometric errors
for the data shown in Table \ref{tbl1}.  First, there is the flux
calibration error of the ACS/HRC instrument, which is 0.5\% 
\citep{sir05}.  Second, there is the uncertainty in performing 
photometry on targets within calibrated, multidrizzled images.  
For targets with no neighbors within $0.''5$, the photometric 
uncertainty is essentially the inverse of the S/N, which is $\leq$
4\% for all 75 photometric targets (i.e. all white dwarfs and 
companions), and $\leq$ 2\% for all but 9 sources.  Third, there is
the error introduced by transforming the flux in the F814W bandpass
to the Cousins $I$ band, which according to \citet{sir05} is no
more than a few percent.  Fourth, there is the error introduced
when photometry was executed on a target with a close ($<0.''5$)
companion.  After PSF subtraction, a few companion point sources
retained a very small amount of residual flux which fell into the
$r=2-3$ pixel photometric aperture of the target star.  This is
smaller than $2-3$\% of the subtracted target's peak flux (\S2.4).
Fifth, there is error introduced by using photometric apertures
of $r=2-3$ pixels which were then corrected to $r=5$ pixels.
The standard deviations in these correction factors were $1-3$\%, 
depending on aperture size and spectral type.  The end result
of adding all of these photometric errors in quadrature is that: 
typical errors for isolated point sources are $<$ 5\%, while those
for point sources with close companions are typically $<$ 6\%.

Only relative astrometry between multiple system components has
been perfomred in the study.  The astrometric errors for point 
sources without close companions are completely due to the uncertainty 
in centroiding, which is strictly a function of S/N.  These are typically
$\leq$ 0.04 pixels ($0.''0015$ for S/N $\geq$ 50).  For targets with close
companions, the measured centroid can be biased in the direction of
the companion, if its flux is not removed by PSF subtraction.  Although
the measured centroid uncertainty for these binary components had essentially
the same range of errors as for single point sources, it is possible that
small biases remained for targets contaminated by any positive or negative
residual flux from its PSF-substracted close companion.  Using $2-3$ pixel
radii for centroiding and comparing the value obtained for a target with a
close companion before and after PSF subtraction of the companion, shifts
were measured that were no larger than 0.2 pixels ($0.''005$).  The centroiding
errors for components of close binaries are no larger than this shift.

\subsection{Resolved Systems}

As can be seen in Figures $1-5$ and Table \ref{tbl1}, there
are 28 systems for which 2 or more objects were totally or
partially resolved (one of which is a previously known wide
red dwarf companion that fell within the ACS field of view and
was itself resolved into 2 close components).  Generally speaking,
the ACS multidrizzled PSFs at F814W had FWHMs $< 0.''077$ and were
symmetric to within 5\% of unity in the ratios of their major
to minor axes by gaussian profile fits.  This allowed mostly
resolved binaries with clearly separated Airy disks to be imaged
down to $0.''09$ in separation at $\Delta m=1-2$ magnitudes
between the components.  For those binaries separated by $>0.''4$
(17 in all), there was very little or no overlap between component
PSFs, and both the white dwarfs and red dwarfs were used to derive
aperture corrections to be used for more closely separated pairs.

It should be noted that there are a few resolved binaries at
separations $>2''$, which, in principle, would be resolvable
from the ground with good seeing.  The reason for this is that
the sample was selected from unresolved 2MASS point sources,
where the beam size is approximately $2''$ at $J$ band in the
images provided by the archive server.  Typically, equally
luminous binaries at magnitudes $J\approx12-15$ mag can appear
partially resolved in 2MASS at separations of $2-4''$ \citep{wac03}.
However, the fact that most red dwarfs in this sample are $2-3$
magnitudes brighter than their white dwarf primaries at $J$ band
makes it difficult or impossible to identify them as binaries in
the 2MASS images.

\subsection{Partially Resolved Systems}

For imaged binaries separated by $<0.''4$, there are two varieties:
those in which two distinct Airy disks are present and those in which
there is a single, elongated Airy disk.  Pairs with two distinct PSF
cores were treated as described above, while those with elongated cores
proved problematic during PSF fitting and subtraction aimed at revealing
individual point sources.  Relative to the apparently single stars in the
ACS image set, these closest binaries all display significantly larger
than normal residuals after single PSF subtractions are performed,
corroborating their binarity.

There are three systems (0949$+$451, 1419$+$576, 1631$+$781)
whose images show the presence of binaries with likely separations
$<0.''025$, corresponding to a single ACS/HRC pixel.  In all three 
cases, the elongated Airy disk is associated with the red dwarf 
component of a resolved binary; i.e., these systems are all triples.
This is a fortunate situation because in each case the white dwarf 
component can be used as a comparison PSF.  These systems are 
described in \S3.7.

The separations of these very close binary systems were estimated
in the following manner.  All the measured PSFs in Table \ref{tbl1}
show some very minor elongation that is typically on the order of one
milliarcsecond.  Specifically, the difference between the major and
minor axes of the gaussian profile fits is typically $\la1-2$ mas.
For the three close doubles, these differences, or elongations, lie
in the range $7-9$ mas.  Equally luminous components were assumed
and the separation of the binary was taken to be its elongation.
Of course, if the pair is not equally bright, then the separation
will be slightly larger, but contour plots of all three objects
display a high degree of symmetry along the major axes, consistent
with equal luminosity.

Without spatially resolved spectra of the close binary
components of these triple systems, it cannot be firmly
concluded that all three are double red dwarfs.  It is
likely that only one of the systems will be observable
from the ground as a spectroscopic target that is spatially
resolved from its white dwarf primary.  Therefore, it may be
difficult (or impossible) to know with certainty the nature
of the two close components in the remaining two systems.
However, depending on the actual orbital period, it may be
possible with radial velocity monitoring to confirm the nature
of these candidate double red dwarfs from the ground over
a period of one to a few years.  None of the other 24 spatially
resolved white dwarf + red dwarf pairs have separations as
small as the three targets with elongated Airy disks, and it
is a safe assumption (for reasons discussed in \S4) that 
these are double red dwarfs.

\subsection{Unresolved Systems}

There are 15 systems in Table \ref{tbl1} for which there is
strong photometric evidence, and often spectroscopic evidence,
for the presence of an unresolved white dwarf + red dwarf pair.
In fact, a thorough literature search beyond \citet{mcc99}
reveals that a few of these systems are recently (since their
selection as targets for this program) discovered DA+dMe systems,
radial velocity variables, or low mass (He core, $M<0.45$
$M_{\odot}$) white dwarfs -- all of which imply binaries with
separations $\la0.1$ AU \citep{saf93,mar95,sch95,sch96}, consistent
with single point sources in the ACS observations.  In any case,
all of these stars have composite optical and near-infrared colors
perfectly consistent with white dwarf and red dwarf components
\citep{wac03}.
	
For these binaries, it was necessary to calculate the flux
contribution of the white dwarf in order to subtract it and
obtain the flux of the red dwarf companion.  This was done 
utilizing techniques discussed in \citet{far04} and \citet{far05},
and \S3.6, where particular attention was given to avoid biasing
the calculations by excluding white dwarf data which may have
been contaminated by the cool companion star.

There were 6 targets (\S3.7) for which the evidence of binarity was
based on low S/N ($<7$ or $<5$) $K_s$ data in the 2MASS point
source catalog.  These were included in the HST/ACS survey as
low priority targets and do not belong to the sample of highly
probable white dwarf + red dwarf binaries.  As yet, none
of these targets have revealed companions in the ACS observations.

\subsection{Establishing Physical Multiplicity}

Any study of stellar multiplicity must address the
likelihood of physical association between putative
companions.  There are several reasons why all the targets
designated here as multiples have a very high probability
of being gravitationally bound.  The first and foremost is
spatial proximity -- the white dwarf targets are associated
with a single point source in 2MASS images.  Second, the
combination of optical and near-infrared colors yields
photometric distance ranges for each component that overlap
when appropriate, mundane white dwarf and red dwarf stellar 
parameters are assumed.  Third, some of the systems imaged
in this study have published references listed in Tables
$1-4$ containing spectroscopic confirmation of their red dwarf
companions in unresolved, composite observations, and/or
previously established common proper motion.  Fourth,
common proper motion is almost certain for all posited
multiples over the timescales since the identification
of the white dwarf component $20-50$ years ago; otherwise
any false pairs should separate when ``blinking'' images
available through the Digitized Sky Survey (e.g., in the
northern hemisphere, the first and second epoch Palomar
Observatory Sky Survey plates).  Fifth, there exist several
white dwarf + red dwarf studies which have established
spectroscopic and/or common proper motion confirmation for
numerous targets of a nearly identical nature \citep{sch96,
ray03,far04,far05}.  Therefore, the multiple stellar systems
presented here should be regarded as physically associated
until shown otherwise.

\subsection{Spectral Type Constraints \& Component Identification}

To determine the projected companion separations in
astronomical units, distances to each binary had to be assessed.
Table \ref{tbl2} lists the stellar parameters from the literature
for the white dwarf primaries in Table \ref{tbl1}.  The first column
contains the white dwarf number, followed by effective temperature,
surface gravity, apparent visual magnitude, photometric distance, 
and references.  A single digit following the decimal place for log
$g$ indicates an assumption of $M=0.60$ $M_{\odot}$.  The $V$
magnitude is either a value uncontaminated by the red dwarf or 
one derived (based on effective temperature and models, magnitudes
and colors in other filters, or photographic magnitudes and colors;
\citealt{far04,far05}) to more correctly reflect the likely 
uncontaminated value.  Distances were calculated from absolute
magnitudes for individual white dwarfs, using the models of 
\citet{ber95a,ber95b} as well as specified references in Table
\ref{tbl2}.

Less than half of the white dwarf targets in Table
\ref{tbl2} have well-determined stellar parameters
(i.e., $T_{\rm eff}$, log $g$) in the literature and,
hence, fairly reliable distance estimates.  For these
well-studied white dwarfs there are published 
parameters based on data (typically spectra, but also
photometry where available) and analyses which are 
not contaminated or biased by the light from their
red dwarf companions.  Fortunately, most if not all
ground-based white dwarf studies are performed in 
the $3000-6000$ \AA \ range, where most red dwarfs
should not contribute significantly.  However, there 
are plenty of cases where flux is seen at these
wavelengths and this has been taken into account
in Table \ref{tbl2} and often by authors in
the corresponding references.

For the remainder of the targets, published white dwarf
parameters do not exist at present; McCook \& Sion (1999
and references therein) sometimes contain $UBV$ or other
optical photometry, but frequently there is only a single
photographic magnitude and/or a spectral type with no
temperature index.  Therefore, at worst, the values
in Table \ref{tbl2} represent conservative best
guesses -- assuming parameters which would make 
the white dwarf and any companion typical -- but
more often they are guesses informed by available data.
Some of the ways in which the effective temperatures and
visual magnitudes were estimated for white dwarf targets
include: (1) published $U-B$ color, which is unlikely to
be contaminated by a cool companion; (2) the implied
color indices (e.g., $B-I$) for any white dwarf which
was resolved from its companion in the ACS observations;
(3) information on the colors and spectral type of the
companion star from 2MASS data, ACS photometry, or other
literature sources; (4) any published optical spectrum 
revealing one or both of the binary components (but without
parameter determinations).

Generally, for each target all available information
was gleaned from the USNO-B1.0 \citep{mon03}, 2MASS All-Sky
Point Source \citep{cut03}, and SuperCOSMOS Sky Survey catalogs
\citep{ham01} to assist in constraining and disentangling binary
component stellar parameters.  Although photographic photometry
can have large absolute calibration errors, typically $\sim0.3$
mag, resultant colors tend to be as accurate as $\la0.1$ mag
for the magnitude range spanned by the white dwarf targets here
\citep{ham01} and can be useful in a number of ways.  Any
colors assist in the assessment of where the flux of the red
dwarf begins to dominate the spectral energy distribution of
the binary, which is typically shortward of 8000 \AA.  If
published optical photometry or spectra were suspected of 
contamination (i.e., effective temperature underestimated in the
literature) then adjustments using model colors based on a more 
likely (conservative) temperature were made.  In this way, the
best available data and estimates were used for all targets in
order to constrain white dwarf parameters and subsequently
deconvolve any composite $IJHK$ data to obtain colors and
magnitudes for red dwarf companions, following methods described
fully in \citet{far04} and \citet{far05}.  Where model white
dwarf parameters were needed, they were taken from \citet{ber95a,
ber95b} and P. Bergeron (2002, private communication).  Where
empirical red dwarf parameters were needed, they were taken
from \citet{kir94} and \citet{dah02}.

A separate but related issue is the identification of
the hot and cool components for the numerous binaries
resolved in the ACS observations.  Utilizing techniques
discussed above, and because in almost every case there
was a significant brightness difference ($\Delta m\ga1$ mag)
between the components, there is high confidence that the
white dwarf component was correctly identified in the ACS 
images (possible exceptions are discussed in \S3.7). Specifically,
the implied $I-K$ color of the red dwarf and the $V-I$ color
of the white dwarf are consistent only if the components were
correctly identified; reversing identities leads to contradictions.
Additionally, the ACS camera scatters light at red wavelengths,
yielding significant haloes for all bright objects (see Figure
5 of \citealt{sir05}), and broadening the PSF of any red
objects relative to blue objects \citep{sir05}.  The measured
PSF sizes in Table \ref{tbl1} reflect this phenomenon, with a
single possible exception discussed in \S3.7.  Despite apparent
good agreement and reassurance that the individual components have
been correctly identified, it is by no means absolute.  In a few
cases, it is possible that an apparently single, resolved binary
component is itself an unresolved double red dwarf binary or
white dwarf + red dwarf binary.

Tables \ref{tbl3} and \ref{tbl4} list the measured
and derived parameters of the red dwarf companion stars.
In the first column is the white dwarf number followed
by the companion name.  The third column contains the spectral
type estimate based on the $I-K$ color listed in the fourth
column.  The fifth column lists the photometric distance to the
red dwarf based on empirical M dwarf data \citep{kir94,dah02}.
It should be mentioned that the white dwarf distance estimates
listed in Table \ref{tbl2} do not always agree with
the implied distance to the red dwarf from absolute
magnitude-spectral type relations \citep{far04,far05}.  There
are several reasons why this might happen: (1) the white dwarf
has a mass significantly above or below the typically assumed
0.60 $M_{\odot}$ value for field DA stars; (2) the effective
temperature of the white dwarf has been poorly estimated; (3)
the spectral type of the red dwarf has been poorly estimated;
(4) either the white dwarf or red dwarf is itself an unresolved
binary; (5) the intrinsic width of the lower main sequence in a
Hertzsprug-Russell (or reduced proper motion) diagram is at least
$\pm1$ magnitude, the slope is steep, and placement of a component 
depends upon both metallicity and age.  This alone can impose on
the order of $1-2$ magnitudes of apparent discrepancy with the
white dwarf estimate when photometric distances are all that
is available.  These are some of the reasons why spectral type
estimation was based on color, not on any absolute magnitude 
implied by the (often nominal) photometric distance to the white
dwarf \citep{far04,far05}.

In principle, obtaining a good photometric distance to nearly
all the white dwarfs in the sample is feasible, since the bulk
are DA white dwarfs which can be spectroscopically fit sans
contamination (by avoidance if necessary) in the $3000-5000$
\AA \ region for $T_{\rm eff}$ and log $g$.  However, a similar
spectroscopic assessment is not possible for the red dwarf
secondaries; spectral types can be determined to limited
accuracy by subtracting the expected white dwarf contribution
over $6000-10000$ \AA, but the equivalent of a stellar radius
(log $g$) determination does not exist. Using empirical
absolute magnitude-spectral type relations is likely to prove
reliable in most cases (assuming knowledge of an accurate spectral
type); however, there is intrinsic scatter in absolute magnitudes
for a given red dwarf spectral type, potentially compounded up
to $\sim1-2$ magnitudes by metallicity and/or multiplicity
\citep{giz97,rei00,far04,far05}.  For comparison and
completeness, the red dwarf distance from color-magnitude
relations is also listed in the tables.

\subsection{Notes on Individual Objects}

{\em 0023$+$388} is more likely closer to $d\approx60$ pc
than 24 pc as suggested by references in \citet{mcc99}.  The
distance underestimate is likely due to the unresolved red
dwarf causing the white dwarf to appear redder, cooler and,
hence, nearer \citep{far04}.

{\em 0131$-$163} is a singular case in which it was difficult
to distinguish which resolved star is the white dwarf or red
dwarf.  The parameters in Table \ref{tbl3} present the most
consistent scenario, but a full optical spectrum and multiband
photometry should be able to reveal which star contributes
more flux around 8000 \AA.

{\em 0208$-$153} is the only resolved pair in which the PSF
widths do not follow the pattern of $\theta_{rd}>\theta_{wd}$.
The reason for this is unclear, but given the fact that ACS 
scatters more light for redder objects, this white dwarf
could conceivably harbor an unresolved red companion.

{\em 0237$+$115, 0347$-$137, 1218$+$497, 1333$+$005, {\rm and} 1458$+$171}
all appear to have discrepancies betwen the distances implied by the
(sometimes inferred) brightness of the white dwarf and red dwarf binary
components.  Many of the white dwarfs have reliable log $g$ determinations
\citep{dre96,koe01,lie05} but the cool companion appears too bright for
its color and, hence, may be a binary.  In other cases, the white dwarf
may be overluminous (or the red dwarf may be underluminous).  These cases
are those that stand out presently, but spectroscopic observations (which
are currently being obtained) may resolve them.

{\em 0324$+$738} has many measurements in \citet{mcc99} and
the current online version of that catalog that are either
contaminated by nearby background stars or were performed on
the wrong stars.  Recently published values of $T_{\rm eff}=4650$
K and an extremely low mass \citep{ber01} are also likely due to the
same phenomenon, as the white dwarf is currently moving between two
background stars and has a nearby ($a\approx13''$) red dwarf common
proper motion companion (which is incorrectly identified as the
white dwarf by coordinates in \citealt{mcc99} and finder charts
provided in the current online version of that catalog).  These
four stars are all currently within a $7''$ radius of one another.
For the present work, the correct stars were identified by using
Digitized Sky Survey images to confirm the pair by their common
proper motion over 40 years and by their photographic magnitudes,
colors and positions in the USNO-A2.0, USNO-B1.0, and 2MASS catalogs
\citep{mon98,mon03,cut03}.  From the ACS images, the white dwarf
has coordinates of $03^{\rm h} 30^{\rm m} 13.89^{\rm s}, +74\arcdeg
01' 57.''1$ while the brighter component of the double red dwarf
(resolved in the observations) is located at $03^{\rm h} 30^{\rm m}
14.37^{\rm s}, +74\arcdeg 02' 09.''7$, both epoch 2005.09.  Using
\citet{gre86} and the ACS data gives $V-I\approx 0.24$ for the white
dwarf and (if correct) together with its measured parallax of
$\pi=0.''025$ would give $T_{\rm eff}\approx9000$ K and log
$g\approx8.74$ ($M\approx1.05$ $M_{\odot}$).  The properly 
identified white dwarf does not have any published, accurate,
and uncontaminated near-infrared data and, hence, there is
currently no evidence for any close (i.e. unresolved by ACS),
cool companion.

{\em 1133$+$489} is the same object as SDSS J113609.59+484318.9
for which \citet{van05} recently determined that the white dwarf
is a DB star with $T_{\rm eff}>38,000$ K, despite the presence of He
II absorption at 4686 \AA \ in their spectrum, together with an M6
or later companion.  The white dwarf was found to be a composite
white dwarf by \citet{gre86} and \citet{wes85} who correctly
determined the primary to be a type DO star with $T_{\rm eff}=47,500$
K plus a cool companion.  However, a companion as late as M6 would
place this system at a distance of about 166 pc based on $JHK_s$ from
2MASS, implying $M_V=11$ mag for the white dwarf and a mass $>1.2$
$M_{\odot}$.  The present analysis finds the $I-K$ color of the
companion gives a spectral type of M5, consistent if the white
dwarf has log $g\approx8.3$ ($M\approx0.84$ $M_{\odot}$).

{\em 1247$+$550, 2211$+$372, 2323$+$256, {\rm and} 2349$-$283} are all
low priority targets whose 2MASS photometry in the Second Incremental
Data Release, despite low S/N at $K_s$, suggested near-infrared excess
\citep{wac03}.  Some of these targets have $JHK_s$ data in the All
Sky Point Source Catalog which differ significantly from the preceeding
catalog values \citep{cut03}.  None of these white dwarfs show any
evidence of companions in the ACS images, and until more accurate
near-infrared photometry is available, binarity should be considered
unlikely.

{\em 1517$+$502} is a rare DA+dC (white dwarf + dwarf carbon
star) system \citep{lie94} which remained unresolved in the ACS
observation.  The expected magnitude difference at $I$ band is 
1.05 mag; thus, conservatively speaking, a separation of
$\ga0.''05$ ($\approx20$ AU) can likely be ruled out.

{\em 1603$+$125} is listed in \citet{mcc99} as a possible magnetic
white dwarf, type DAH.  Based on the deconvolved, $I-K=1.38$, color for
the companion, a spectral type of K3 is inferred.  This places the system
at $d\approx1660$ pc and implies that the primary is very likely an
sdB star with $M_V\approx4.5$ mag \citep{max00}.

{\em 1619$+$525} displays three objects within a $3''$ radius on
the ACS images; their separations and magnitudes are listed in Table
\ref{tbl1}.  The photometry was performed with color and aperture
corrections appropriate for one white dwarf and two red dwarfs of
the implied intrinsic faintness and color.  If this is a physical
triple, then the ratio of the projected separations is 5.6 to 1.
The companionship of the two brighter stars is quite firm as the
elongated pair can be seen on at least 2 Digitized Sky Survey
images, almost certainly comoving over 43 years.  The third
candidate companion lies within $0.''5$ of the white dwarf, which
is located at $|b|=43.9\arcdeg$.  Hence, they are quite likely
to be physically associated.  Additionally, the candidate third 
component appears quite red, as would be expected of a late M dwarf;
the 2MASS images are elongated at $JH$, and especially at $K_s$, along
the correct position angle.  This is consistent if the tertiary candidate
is a companion because, although the brightness difference of the two
red dwarfs is 2.8 magnitudes at $I$ band, it should become $<2$
magnitudes in the near-infrared if the two contributing components
are around M5 and DA2.8+M7 \citep{kir94,dah02}.  Attempts to deconvolve
the 2MASS magnitudes into two components failed, however, and the 2MASS
photometry may be inaccurate due to the presence of two objects (the
white dwarf should contribute very little in the near-infrared).
Ground-based followup should conclusively demonstrate that this
system is a DA2.8+dM5+dM7 (approximate M types) triple system.

{\em 1631$+$781} is a triple system with a well-resolved red
dwarf companion which is itself a barely resolved double.  This
system is mentioned in 47 papers in the literature since its
discovery by \citet{coo92}, at least 12 of which mention or
discuss its nature as a precatacylsmic variable or post-common
envelope binary.  However, both \citet{sio95} and \citet{sch96}
demonstrated the unlikelihood that the main binary lies in a
close orbit by finding an absence of radial velocity variability
in the Balmer emission lines from the companion.  \citet{cat95}
added the possibility that the system is face on or has low
enough inclination to preclude significant variations in radial
velocity as seen from Earth.  \citet{ble02} conclude that the
system is not face on from a positive detection of rotational
broadening ($v_{rot}\sin{i}=25.2\pm2.3$ km s$^{-1}$) in the red
dwarf at several lines in a high resolution optical spectrum.
Given that all components of the triple system are seen in the
ACS image, it is entirely possible that this system has a low
inclination, and the line broadening reported by \citet{ble02}
might be due in part or in whole to the binarity of the red dwarf
component.  Somewhat surprisingly, \citet{mor05} lists this system
among all known detached, post-common envelope binaries with known
periods, citing \citet{fuh03} who report x-ray variability with
a period of $69.4\pm8.3$ hours and a false-alarm probability of
4.4\%.  Since the main binary is quite widely separated, the
source of the x-ray variations are either intrinsic to the
hot degenerate star itself (i.e., perhaps a starspot) as seen in
other single white dwarfs \citep{fuh03}, or flare activity from
one or both components of the red dwarf pair \citep{rei00}.  In
any case, two things are clear from the ACS data.  First, this
system never shared a common envelope, as the current projected
separation of the double red dwarf from the primary is 17 AU
($P\geq57$ years).  Second, all the documented optical emission
features arise from activity within the double red dwarf system,
either intrinsic to one or both of the stars (interactions are
unlikely at $a\ga0.4$ AU, $P\approx100$ days).
 
{\em 1845$+$683} is a hot white dwarf ($V\approx15.5$
mag, $T_{\rm eff}\approx37,000$ K; see references in Table
\ref{tbl2}) reported as a binary in both \citet{gre00} and 
\citet{hol03}.  It is likely that the second reference is
merely pointing to the work presented in the first reference,
as there exists no other discoverable discussion of suspected 
binarity of this white dwarf in the literature.  The 2MASS
catalog gives $J=16.07\pm0.09$ mag, $H=16.28\pm0.22$ mag, and
an upper limit of $K_s>15.29$ mag for the white dwarf, which
are consistent with a single star of the appropriate effective
temperature, and also consistent with the ACS $I$ band magnitude.
\citet{gre00} have $J=14.85\pm0.10$ mag and $K=14.37\pm0.19$ mag
which apparently corresponds to a field star located $46''$ away at
${\rm PA}=225\arcdeg$, which has  2MASS photometry of $J=14.93\pm0.04$
mag, $H=14.36\pm0.04$ mag, and $K_s=14.30\pm0.09$ mag, all with
S/N $>12$.  All this suggests that the white dwarf is neither a
suspected nor confirmed binary.

{\em 2151$-$015} contains a resolved M8 dwarf, the coolest
companion resolved in the survey so far.  The identity of the
white dwarf and red dwarf components are firm from ground-based
$BVRI$ images, where the M star is seen partially resolved only
at $I$ \citep{far04,far05}.  However, the white dwarf exhibits a
significant halo in the ACS image which is unexpected for a
$\sim8500$ K object, and not seen in other white dwarf primaries
-- even those with unresolved, spectroscopically confirmed,
late M dwarf companions such as 0354$+$463 (Rubin 80, DA6+dM7;
\citealt{far04,far05}).  While this may simply be the result
of the fact that the primary is relatively bright compared
to the other resolved white dwarfs, it might be due to a
third, unresolved, even cooler component.

\section{CONCLUSIONS}

This HST/ACS survey was specifically designed to
search for white dwarf + red dwarf binaries separated by
around one to a few AU.  Theory predicts, and extant observations
support, a bimodal distribution of orbital separations in which low
mass companions vacate this region during the post-main sequence 
evolution of the white dwarf progenitor \citep{jea24,bon85,zuc87,
val88,bon90,dek93,yun93,sch96,liv96,max98,sch03,far04}.  Specifically, 
companions within a few AU should interact both directly and tidally 
with the AGB slow wind and expanding photosphere, imparting some of 
their angular momentum to the envelope and arriving eventually at 
closer orbital semimajor axes.  Those companions originally outside 
of a few AU should eschew the AGB envelope and experience an expansion
of their orbital semimajor axes by a factor proportional to the total 
amount of mass lost   There is no a priori reason to believe
that the phase space around $\sim0.5-5$ AU should be utterly
devoid of low mass main sequence companions to white dwarfs.
In fact, there should be some real width in both peaks of
the actual bimodal distribution of separations, each with
tails overlapping the ``forbidden'' range of semimajor axes.

Figure \ref{fig6} displays the sensitivity of the survey
to companions in the few AU range.  Here, two assumptions
are used; 1) $0.''010$ sensitivity, for companions at $\Delta
m\la0.5$ mag and 2) $0.''025$ sensitivity, for companions at
$\Delta m\la2$ mag.  These assumptions are conservative,
especially given the performed PSF subtractions.  All
targetted binary candidates are expected to have flux 
ratios corresponding $\Delta m<2.5$ mag in the F814W 
filter.  Figure \ref{fig7} shows the distribution of
projected separations for detected companions and upper
limits for unresolved binaries.

So far, no white dwarf + red dwarf systems have been resolved
at separations around a few AU.  The ability of this survey 
to detect binaries of comparable luminosity at such separations
is demonstrated by both Figures \ref{fig6} and \ref{fig7}, plus
the three very close binaries (likely double red dwarfs) in Table
\ref{tbl5}, implying that pairs separated by as little as 1 AU at
$d=100$ pc were at least partially resolvable.  Although accurate
distances are needed for the sample stars to confirm these preliminary
findings, there are not yet any ambiguous cases where a reasonable change
in the distance to the binary would bring the separation into the few
AU range.

If these initial conclusions are correct, these data are the
first empirical evidence for the bimodal distribution of low
mass, unevolved companions to white dwarfs.  This would also
imply that as many as 100\% of the 15 unresolved white dwarf
+ red dwarf pairs are in close orbits, and good candidates
for radial velocity variables.  If the first half of this
survey is representative of the whole, it should be expected
that this program will eventually resolve around 55 total
wide binaries, and identify around 30 total binaries that
are candidate radial velocity variables.

\section{FUTURE WORK}

Followup work is currently being carried out for all 
program stars for which a good photometric distance
determination (i.e., reliable $UBV$ photometry, $T_{\rm eff}$,
and log $g$) does not exist in the literature.  In order to
assess the physical separations of the low mass companions
in AU, an accurate distance is required.  Combined with the
near-infrared data, optical photometry and spectroscopy should 
allow a complete determination of stellar parameters for these
binary and triple systems.

Stellar parameters are not only needed to determine distance
and separation for these binaries, but also to study the system
components themselves.  Another goal of this survey is to compare
the masses (via spectral types) of the red dwarf companions in wide,
resolved systems versus those in unresolved, likely post-common
envelope systems.  Given enough stars in both categories, a
statistical analysis can be made of the resulting spectral
type (and, by proxy, mass) distributions to see what effect,
if any, common envelope evolution has had on the secondary
masses \citep{far04,far05}

\acknowledgments

J. Farihi would like to thank the following people for
their much appreciated software and computing support
directly related to this work: R. Hook, C. Hanley, D.
Starr, K. Labrie, \& K. Chiboucas.  Sincere thanks goes
to the anonymous refereee for many helpful comments which 
improved the quality of the manuscript.  This work is based
on observations made with the Hubble Space Telescope 
which is operated by the Association of  Universities
for Research in Astronomy under NASA contract NAS 5-26555.
Support for Program number 10255 was provided by NASA through
grant HST-GO-10255 from the Space Telescope Science Institute.
Some data used in this paper are part of the Two Micron All
Sky Survey, a joint project of the University of Massachusetts
and the Infrared Processing and Analysis Center (IPAC)/CIT, 
funded by NASA and the National Science Foundation.  2MASS
data were retrieved from the NASA/IPAC Infrared Science 
Archive, which is operated by the Jet Propulsion
Laboratory, CIT, under contract with NASA.  This
research has made use of NASA's Astrophysics Data
System Bibliographic Services, and the SIMBAD
database, operated at CDS, Strasbourg, France.

{\em Facility:} \facility{HST (ACS)}

\clearpage

\begin{figure}
\figurenum{1a}
\epsscale{0.7}
\plotone{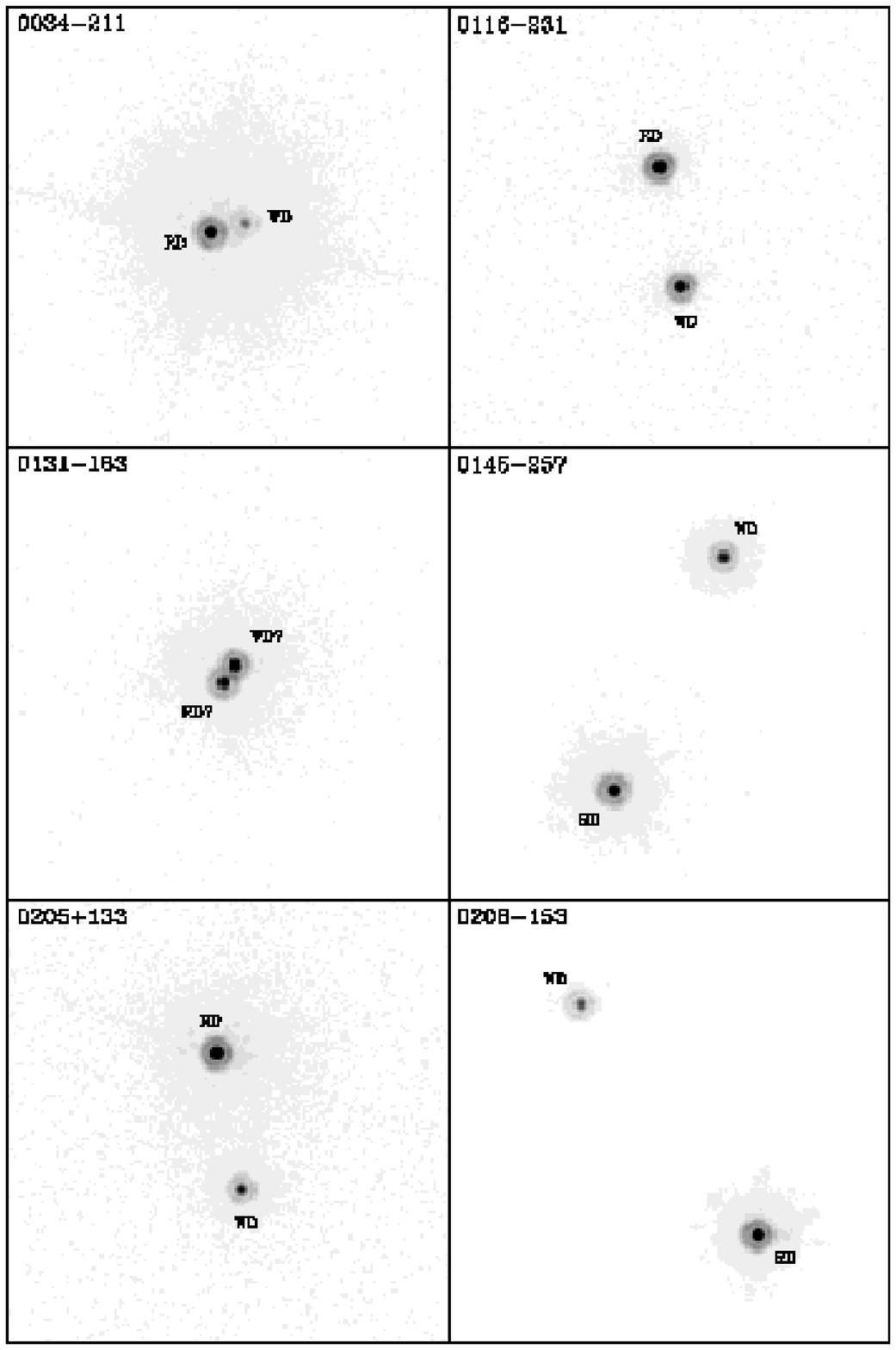}
\caption{Multidrizzled images of the ACS/HRC data taken
in the F814W filter of all 28 totally or partially resolved
multiple systems, plus an example of a single unresolved
point source (0303$-$007).  The images are $4''\times4''$
($0.''025$ pixels) with North up and East left.
\label{fig1}}
\end{figure}

\clearpage

\begin{figure}
\figurenum{1b}
\plotone{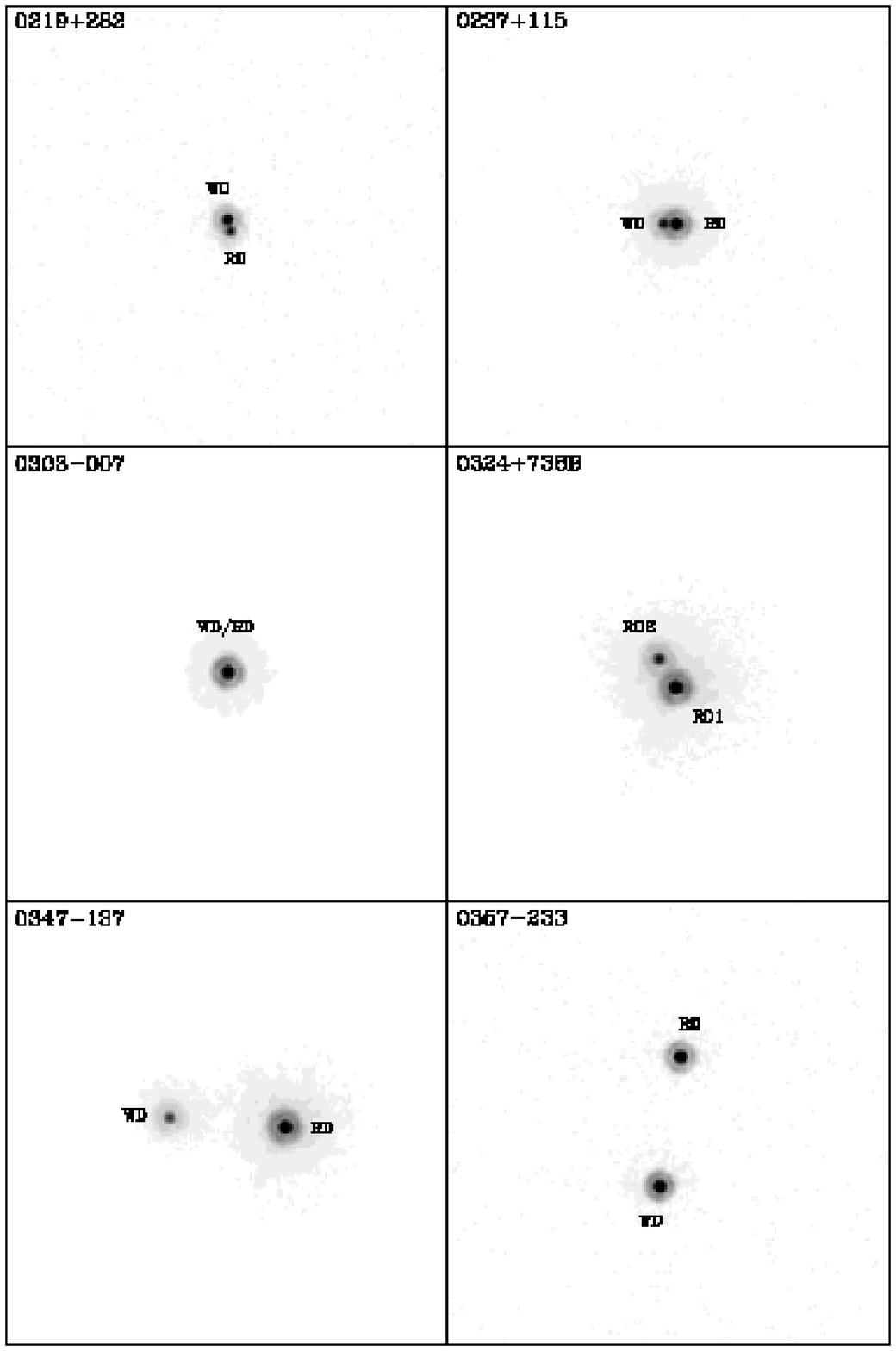}
\caption{see Figure \ref{fig1}.
\label{fig2}}
\end{figure}

\clearpage

\begin{figure}
\figurenum{1c}
\plotone{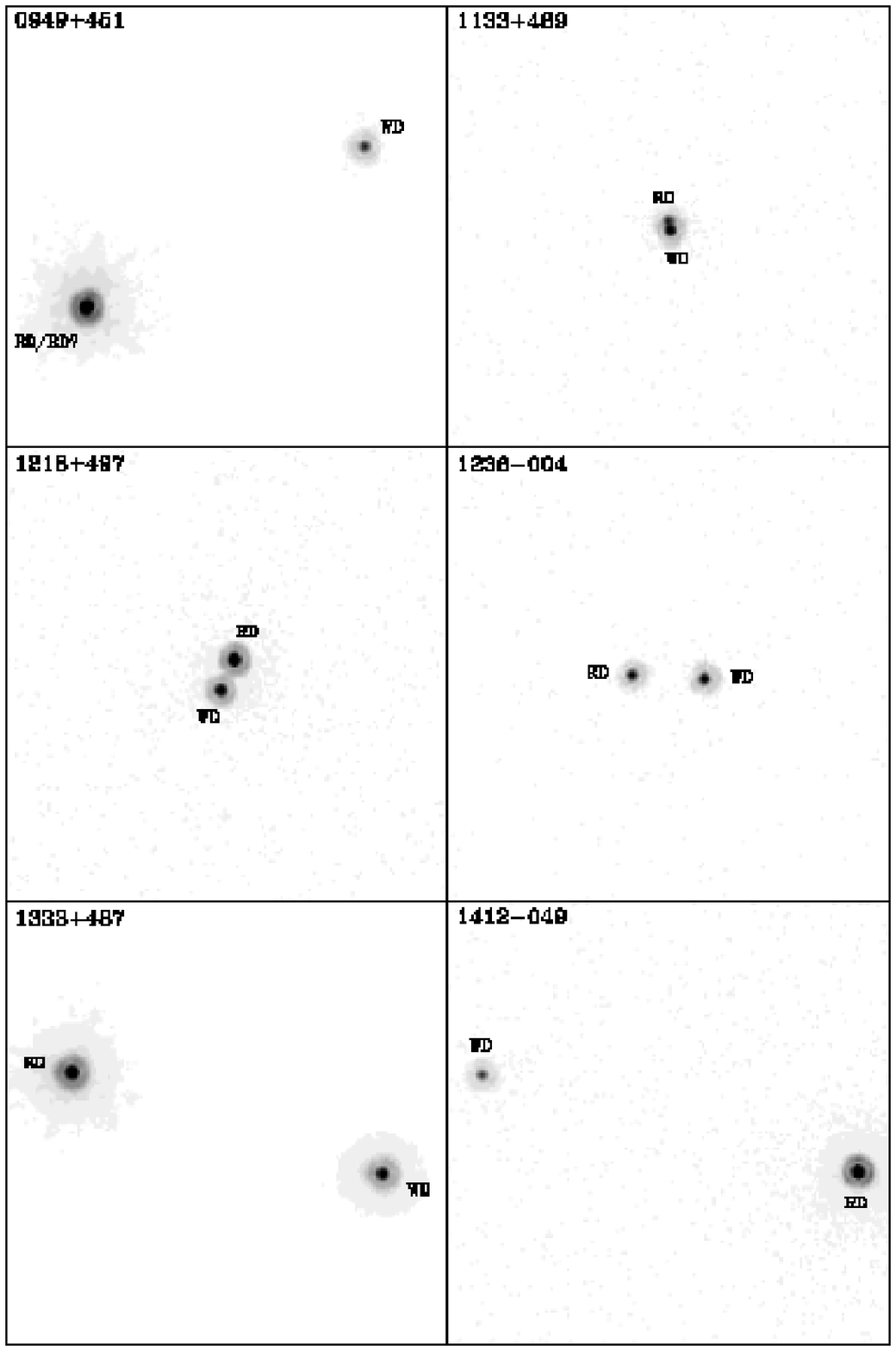}
\caption{see Figure \ref{fig1}.
\label{fig3}}
\end{figure}

\clearpage

\begin{figure}
\figurenum{1d}
\plotone{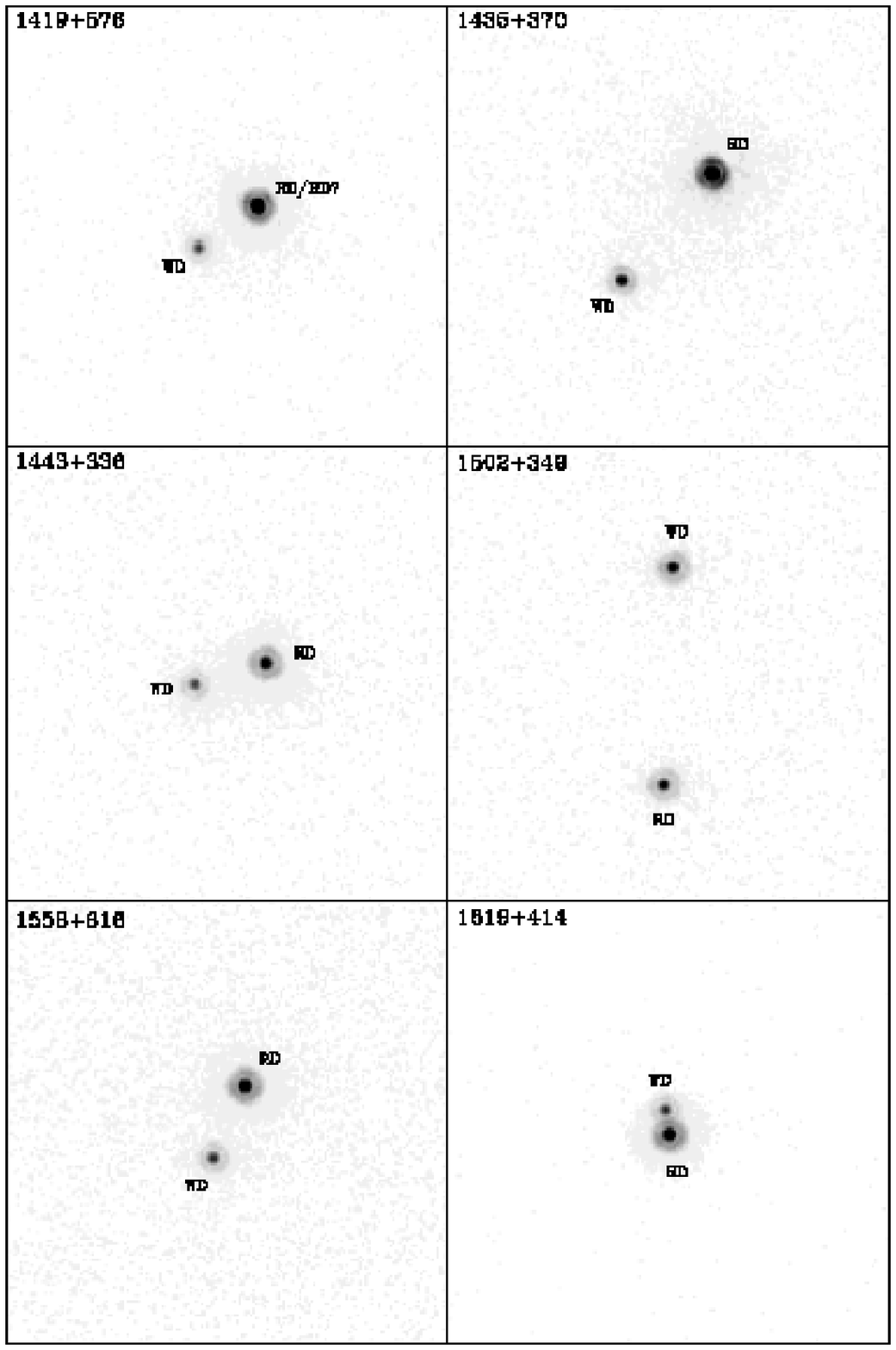}
\caption{see Figure \ref{fig1}.
\label{fig4}}
\end{figure}

\clearpage

\begin{figure}
\figurenum{1e}
\plotone{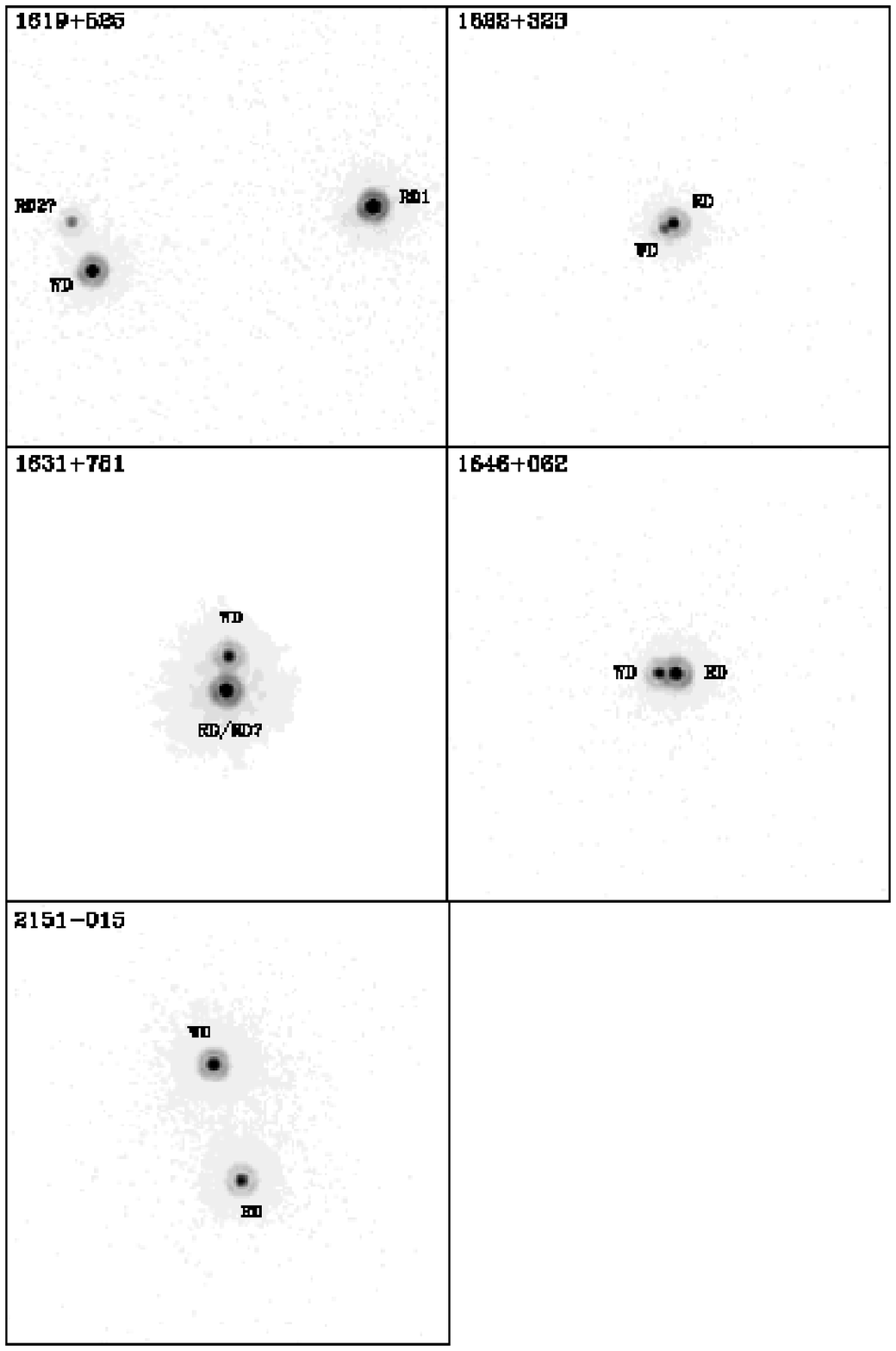}
\caption{see Figure \ref{fig1}.
\label{fig5}}
\end{figure}

\clearpage

\begin{figure}
\figurenum{2}
\epsscale{1.0}
\plotone{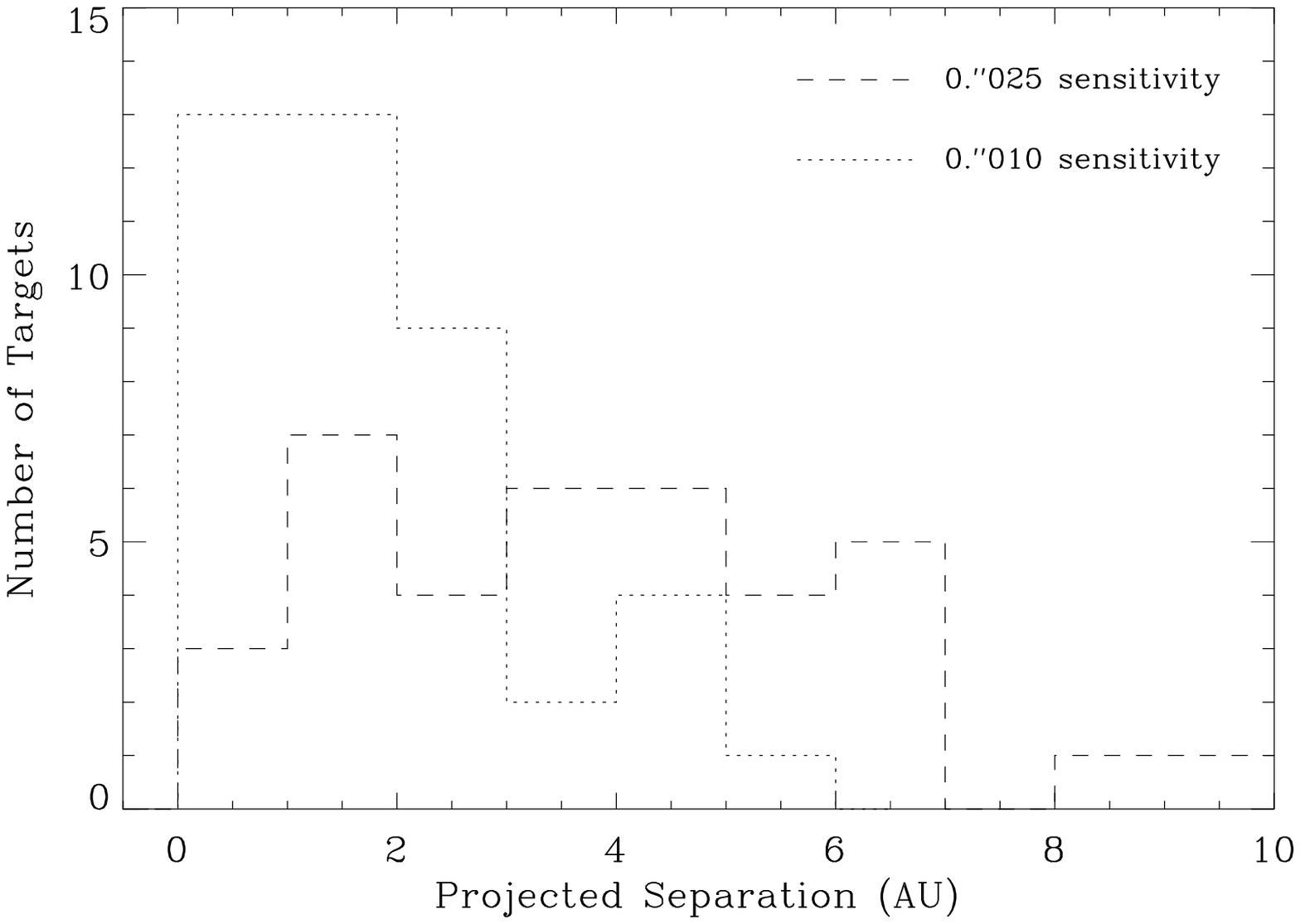}
\caption{The number of targets from Tables \ref{tbl3} and \ref{tbl4}
for which the ACS imaging data were sensitive to companion detection
in the 0.3 AU (the minimum detectable projected separation) to 10 AU 
range.  The bin size is 1 AU with integer bounds.  The dashed line
assumes a sensitivity of $0.''025$ while the dotted line assumes
$0.''010$.  All 41 bona fide white dwarf targets with strong evidence
for near-infrared excess were sensitive to companion imaging
detection within 5 AU, plus numerous targets sensitive to
detections within 1 AU.
\label{fig6}}
\end{figure}

\clearpage

\begin{figure}
\figurenum{3}
\plotone{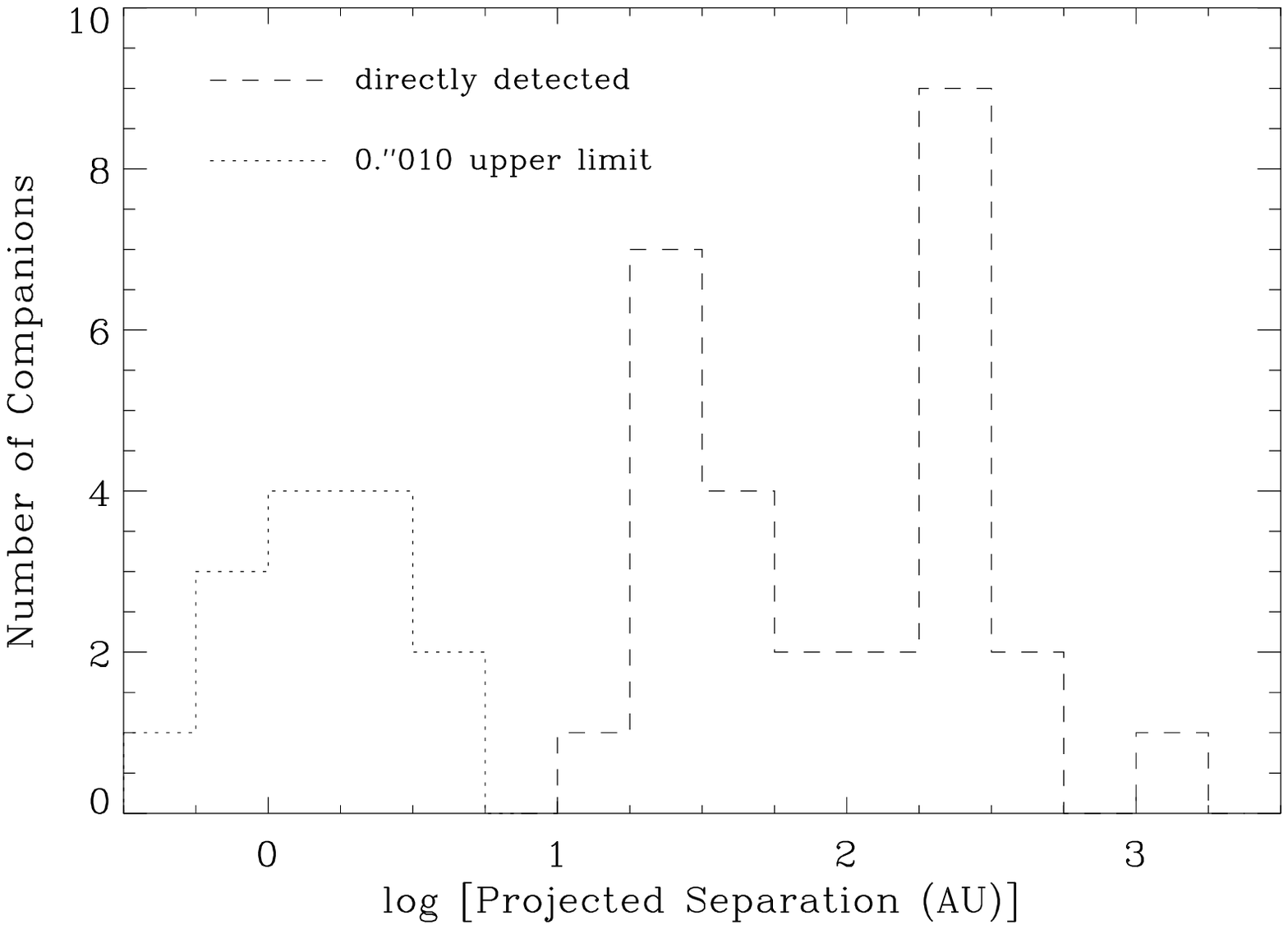}
\caption{The distribution of projected separations for
targets from Tables \ref{tbl3} and \ref{tbl4}.  The bin size
is 0.25 in logarithmic units of AU, with identical bounds.
The dashed line represents directly detected companions,
while the dotted line represents upper limits of $0.''010$
for those binaries which remained unresolved.
\label{fig7}}
\end{figure}

\clearpage

\begin{deluxetable}{llcccccccc}
\rotate
\tablecaption{Target ACS Data\label{tbl1}}
\tabletypesize{\small}
\tablewidth{0pt}
\tablehead{
\colhead{WD\#}			&
\colhead{Name}			&
\colhead{Resolved?}		&
\colhead{FWHM ($''$)}		&
\colhead{$a/b$}			&
\colhead{$a_{sky}$ ($''$)}	&
\colhead{PA ($\arcdeg$)}	&
\colhead{F814W (mag)\tablenotemark{\dag}}		&
\colhead{$I_c$ (mag)\tablenotemark{\dag}}		&
\colhead{Notes}}

\startdata

0023$+$388	&G171-B10A	&No	&0.0747	&1.030	&\nodata	&\nodata	&$15.18\pm0.02$	&$15.18\pm0.03$	&1,6,7\\	

0034$-$211	&LTT 0329	&Yes	&0.0732	&1.034	&\nodata	&\nodata	&$15.10\pm0.04$	&$15.07\pm0.05$	&2,5\\
		&		&	&0.0757	&1.022	&$0.328\pm0.002$&$106.04\pm0.13$&$12.83\pm0.02$	&$12.89\pm0.03$	&5\\

0116$-$231	&GD 695		&Yes	&0.0733	&1.006	&\nodata	&\nodata	&$16.60\pm0.02$	&$16.57\pm0.03$	&2\\
		&		&	&0.0754	&1.019	&$1.105\pm0.002$&$9.90\pm0.04$	&$16.16\pm0.02$	&$16.27\pm0.03$	&\\

0131$-$163	&GD 984		&Yes	&0.0745	&1.017	&\nodata	&\nodata	&$14.28\pm0.03$	&$14.23\pm0.04$	&2,5\\
		&		&	&0.0751	&1.017	&$0.189\pm0.002$&$145.26\pm0.23$&$14.49\pm0.03$	&$14.57\pm0.04$	&5\\

0145$-$257	&GD 1401	&Yes	&0.0740	&1.017	&\nodata	&\nodata	&$14.98\pm0.02$	&$14.86\pm0.03$	&2\\
		&		&	&0.0761	&1.011	&$2.295\pm0.002$&$154.11\pm0.02$&$13.83\pm0.02$	&$13.91\pm0.03$	&\\
		
0205$+$133	&PG		&Yes	&0.0737	&1.044	&\nodata	&\nodata	&$15.44\pm0.02$	&$15.27\pm0.03$	&2\\
		&		&	&0.0762	&1.040	&$1.257\pm0.002$&$10.68\pm0.03$	&$13.91\pm0.02$	&$13.97\pm0.03$	&\\ 
						
0208$-$153	&MCT		&Yes	&0.0753	&1.026	&\nodata	&\nodata	&$15.93\pm0.02$	&$15.90\pm0.03$	&2\\ 
		&		&	&0.0741	&1.021	&$2.647\pm0.002$&$217.39\pm0.02$&$13.78\pm0.02$	&$13.83\pm0.03$	&\\ 
		
0219$+$282	&KUV		&Yes	&0.0750	&1.032	&\nodata	&\nodata	&$17.33\pm0.03$	&$17.29\pm0.04$	&2,5,16\\
		&		&	&0.0773	&1.086	&$0.117\pm$0.005&$195.2\pm1.2$	&$18.13\pm0.06$	&$18.29\pm0.07$	&5\\
			
0237$+$115	&PG		&Yes	&0.0756 &1.046	&\nodata	&\nodata	&$16.35\pm0.05$	&$16.31\pm0.06$	&2,5,14\\
		&		&	&0.0760	&1.020	&$0.124\pm0.005$&$268.2\pm1.2$	&$15.01\pm0.03$	&$15.05\pm0.04$	&5\\
						
0303$-$007	&KUV		&No	&0.0758	&1.024	&\nodata	&\nodata	&$14.54\pm0.02$	&$14.57\pm0.03$	&1\\

0324$+$738	&G221-10	&No	&0.0751	&1.042	&\nodata	&\nodata	&$16.61\pm0.02$	&$16.60\pm0.03$	&3,6,7,10\\
		&G221-11	&Yes	&0.0767	&1.028	&\nodata	&\nodata	&$13.68\pm0.02$	&$13.82\pm0.03$	&2,5,19\\
		&		&	&0.0760	&1.037	&$0.297\pm0.003$&$31.48\pm0.29$	&$15.11\pm0.04$	&$15.30\pm0.05$	&5\\ 
			
0347$-$137	&GD 51		&Yes	&0.0745	&1.017	&\nodata	&\nodata	&$15.58\pm0.02$	&$15.55\pm0.03$	&2\\
		&		&	&0.0767	&1.024	&$1.052\pm0.002$&$265.16\pm0.04$&$13.64\pm0.02$	&$13.71\pm0.03$	&\\
		
0354$+$463	&Rubin 80	&No	&0.0742	&1.017	&\nodata	&\nodata	&$14.91\pm0.02$	&$14.90\pm0.03$	&1\\

0357$-$233	&Ton S 392	&Yes	&0.0745	&1.017	&\nodata	&\nodata	&$16.22\pm0.02$	&$16.18\pm0.03$	&2\\
		&		&	&0.0755	&1.014	&$1.190\pm0.002$&$351.00\pm0.04$&$16.46\pm0.02$	&$16.52\pm0.03$	&\\
			
0458$-$662	&WD		&No	&0.0750	&1.021	&\nodata	&\nodata	&$14.56\pm0.02$	&$14.58\pm0.03$	&1,9,11\\

0949$+$451	&HS		&Yes	&0.0744	&1.016	&\nodata	&\nodata	&$15.90\pm0.02$	&$15.89\pm0.03$	&2,6\\ 
		&		&	&0.0782	&1.125	&$2.892\pm0.002$&$119.66\pm0.02$&$13.53\pm0.04$	&$13.58\pm0.05$	&4\\ 
		&		&	&	&	&		&		&$14.3\pm0.2$	&$14.3\pm0.2$	&8\\
		&		&	&	&	&$0.009\pm0.005$&$350\pm16$	&$14.3\pm0.2$	&$14.3\pm0.2$	&8\\	

1051$+$516	&SBS		&No	&0.0755	&1.007	&\nodata	&\nodata	&$14.68\pm0.02$	&$14.70\pm0.03$	&1,9\\

1133$+$489	&PG		&Yes	&0.0740	&1.030	&\nodata	&\nodata	&$17.53\pm0.04$	&$17.49\pm0.05$	&2,5,14\\
		&		&	&0.0756	&1.076	&$0.094\pm0.005$&$10.0\pm1.5$	&$18.18\pm0.06$	&$18.34\pm0.07$	&5\\

1218$+$497	&PG		&Yes	&0.0736	&1.029	&\nodata	&\nodata	&$16.74\pm0.02$	&$16.70\pm0.03$	&2,5\\
		&		&	&0.0739	&1.022	&$0.302\pm0.002$&$335.22\pm0.15$&$16.14\pm0.02$	&$16.21\pm0.03$	&5\\
			
1236$-$004	&WD		&Yes	&0.0747	&1.031	&\nodata	&\nodata	&$17.93\pm0.03$	&$17.89\pm0.04$	&2\\
		&		&	&0.0756	&1.015	&$0.658\pm0.004$&$87.30\pm0.17$&$18.05\pm0.04$	&$18.16\pm0.05$	&\\

1247$+$550	&LHS 342	&No	&0.0756	&1.013	&\nodata	&\nodata	&$16.41\pm0.02$	&$16.43\pm0.03$	&1,7,10,12\\

1333$+$005\tablenotemark{*}
		&LP 618-14	&No	&0.0762	&1.033	&\nodata	&\nodata	&$15.64\pm0.02$	&$15.73\pm0.03$	&1,15\\

1333$+$487	&GD 325		&Yes	&0.0747 &1.020	&\nodata	&\nodata	&$14.24\pm0.02$	&$14.23\pm0.03$	&2,13\\
		&		&	&0.0764	&1.015	&$2.947\pm0.002$&$71.97\pm0.02$	&$13.39\pm0.02$	&$13.50\pm0.03$	&\\
		
1339$+$606	&RE		&No	&0.0738 &1.026	&\nodata	&\nodata	&$16.48\pm0.02$	&$16.48\pm0.03$	&1\\

1412$-$049	&PG		&Yes	&0.0722	&1.040	&\nodata	&\nodata	&$17.10\pm0.03$	&$17.06\pm0.04$	&2\\ 
		&		&	&0.0747	&1.032	&$3.508\pm0.002$&$255.53\pm0.02$&$14.76\pm0.02$	&$14.82\pm0.03$	&\\ 
			
1419$+$576	&SBS		&Yes	&0.0741	&1.019	&\nodata	&\nodata	&$17.54\pm0.03$	&$17.52\pm0.04$	&2,6\\ 
		&		&	&0.0778	&1.114	&$0.658\pm0.002$&$304.06\pm0.15$&$15.05\pm0.04$	&$15.11\pm0.05$	&4\\ 
		&		&	&	&	&		&		&$15.8\pm0.2$	&$15.9\pm0.2$	&8\\
		&		&	&	&	&$0.008\pm0.005$&$205\pm18$	&$15.8\pm0.2$	&$15.9\pm0.2$	&8\\
			
1433$+$538	&GD 337		&No	&0.0741	&1.005	&\nodata	&\nodata	&$15.68\pm0.02$	&$15.69\pm0.03$	&1\\

1435$+$370	&CBS 194	&Yes	&0.0710	&1.074	&\nodata	&\nodata	&$16.93\pm0.02$	&$16.89\pm0.03$	&2\\
		&		&	&0.0754	&1.032	&$1.251\pm0.002$&$318.80\pm0.04$&$14.80\pm0.02$	&$14.85\pm0.03$	&\\		

1443$+$336	&PG		&Yes	&0.0723	&1.033	&\nodata	&\nodata	&$16.93\pm0.02$	&$16.89\pm0.03$	&2\\
		&		&	&0.0745	&1.021	&$0.679\pm0.002$&$286.17\pm0.08$&$15.53\pm0.02$	&$15.60\pm0.03$	&\\ 
			
1458$+$171	&PG		&No	&0.0747	&1.025	&\nodata	&\nodata	&$15.82\pm0.02$	&$15.81\pm0.03$	&1,12\\

1502$+$349	&CBS 223	&Yes	&0.0740	&1.026	&\nodata	&\nodata	&$16.89\pm0.02$	&$16.84\pm0.03$	&2\\
		&		&	&0.0758	&1.030	&$1.913\pm0.002$&$179.31\pm0.02$&$17.07\pm0.03$	&$17.16\pm0.04$	&\\ 

1504$+$546	&CBS 301	&No	&0.0741	&1.039	&\nodata	&\nodata	&$15.02\pm0.02$	&$15.02\pm0.03$	&1,9\\

1517$+$502	&CBS 311	&No	&0.0735	&1.040	&\nodata	&\nodata	&$16.69\pm0.02$	&$16.70\pm0.03$	&1,17\\
				
1558$+$616	&HS		&Yes	&0.0734	&1.023	&\nodata	&\nodata	&$17.29\pm0.03$	&$17.27\pm0.04$	&2\\
		&		&	&0.0768	&1.016	&$0.715\pm0.002$&$336.26\pm0.12$&$15.73\pm0.02$	&$15.82\pm0.03$	&\\
			
1603$+$125	&KUV		&No	&0.0726	&1.044	&\nodata	&\nodata	&$14.15\pm0.02$	&$14.16\pm0.03$	&1,18\\

1619$+$525	&PG		&Yes	&0.0731	&1.013	&\nodata	&\nodata	&$15.81\pm0.02$	&$15.79\pm0.03$	&3,6\\
		&		&	&0.0748	&1.011	&$2.596\pm0.002$&$282.52\pm0.02$&$15.18\pm0.02$	&$15.28\pm0.03$	&\\		
		&		&	&0.0756	&1.040	&$0.466\pm0.003$&$23.97\pm0.16$	&$17.85\pm0.04$	&$18.11\pm0.05$	&\\

1619$+$414	&KUV		&Yes	&0.0735	&1.007	&\nodata	&\nodata	&$17.39\pm0.03$	&$17.36\pm0.04$	&2,5\\
		&		&	&0.0764	&1.008	&$0.231\pm0.002$&$188.79\pm0.38$&$15.67\pm0.02$	&$15.74\pm0.03$	&5\\

1622$+$323	&PG		&Yes	&0.0730	&1.079	&\nodata	&\nodata	&$16.93\pm0.08$	&$16.89\pm0.10$	&2,5\\
		&		&	&0.0747	&1.066	&$0.094\pm0.005$&$300.8\pm1.5$	&$15.78\pm0.04$ &$15.83\pm0.05$ &5\\
			
1631$+$781	&RE		&Yes	&0.0742	&1.011	&\nodata	&\nodata	&$13.62\pm0.03$	&$13.58\pm0.04$	&2,6,9\\
		&		&	&0.0757	&1.102	&$0.302\pm0.002$&$355.87\pm0.14$&$12.29\pm0.04$	&$12.36\pm0.05$	&4\\
		&		&	&	&	&		&		&$13.0\pm0.2$	&$13.1\pm0.2$	&8\\
		&		&	&	&	&$0.007\pm0.005$&$200\pm20$	&$13.0\pm0.2$	&$13.1\pm0.2$	&8\\
			
1646$+$062	&PG		&Yes	&0.0734	&1.025	&\nodata	&\nodata	&$16.46\pm0.04$	&$16.42\pm0.05$	&2,5\\
		&		&	&0.0744	&1.036	&$0.163\pm0.003$&$270.98\pm0.53$&$15.40\pm0.03$	&$15.48\pm0.04$	&5\\

1845$+$683	&KUV		&No	&0.0732	&1.015	&\nodata	&\nodata	&$15.65\pm0.02$	&$15.61\pm0.03$	&1,7\\

2009$+$622	&GD 543		&No	&0.0747	&1.030	&\nodata	&\nodata	&$15.05\pm0.02$	&$15.06\pm0.03$	&1,9,11\\

2151$-$015	&LTT 8747	&Yes	&0.0733	&1.023	&\nodata	&\nodata	&$14.30\pm0.02$	&$14.29\pm0.03$	&2\\
		&		&	&0.0776	&1.024	&$1.082\pm0.002$&$193.73\pm0.04$&$15.07\pm0.02$	&$15.34\pm0.03$	&\\
			
2211$+$372	&LHS 3779	&No	&0.0735	&1.010	&\nodata	&\nodata	&$16.38\pm0.02$	&$16.38\pm0.03$	&1,7,10\\

2237$-$365	&LHS 3841	&No	&0.0750	&1.010	&\nodata	&\nodata	&$14.83\pm0.02$	&$14.93\pm0.03$	&1,15\\

2317$+$268	&KUV		&No	&0.0749	&1.026	&\nodata	&\nodata	&$15.64\pm0.02$	&$15.61\pm0.03$	&1\\

2323$+$256	&G128-62	&No	&0.0739	&1.027	&\nodata	&\nodata	&$16.29\pm0.02$	&$16.29\pm0.03$	&1,7,10\\

2349$-$283	&GD 1617	&No	&0.0734	&1.010	&\nodata	&\nodata	&$15.65\pm0.02$	&$15.62\pm0.03$	&1,7\\
				
\enddata

\tablecomments{
(1) Single point source;
(2) Double point source, two Airy disks;
(3) Triple point source, three Airy disks
(4) Single elongated Airy disk;
(5) Measurements affected by close binarity;
(6) Triple system;
(7) Low S/N in 2MASS;
(8) Equal luminosity assumed;
(9) DA+dMe;	
(10) Cool red WD, $T_{\rm eff}<8000$ K;
(11) Radial velocity variable;
(12) low mass, He core WD;
(13) DB+dM;
(14) DO+dM;
(15) DC+dM;
(16) DBA+dM;
(17) DA+dC;
(18) sdB+dK;	
(19) Common proper motion companion}

\tablenotetext{*}{Not in \citet{mcc99}, WD\# unofficial}

\tablenotetext{\dag}{All photometry is in Vega magnitudes.
Photometric and astrometric errors are discussed in \S3.1.}

\end{deluxetable}

\clearpage

\begin{deluxetable}{lccccc}
\tablecaption{White Dwarf Parameters\label{tbl2}}
\tablewidth{0pt}
\tablehead{
\colhead{WD\#}			&
\colhead{$T_{\rm eff}$ (K)}	&
\colhead{log $g$}		&
\colhead{$V$ (mag)}		&
\colhead{$d_{\rm wd}$ (pc)}		&
\colhead{References}}

\startdata

0023$+$388	&10,400	&8.0	&15.97	&62	&1,13\\

0034$-$211	&17,200	&8.04	&15.03	&63	&1,4,5,14,15\\

0116$-$231	&25,000	&7.9	&16.29	&164	&1,16,17\\	

0131$-$163	&50,000	&7.75	&13.90	&109	&1,3,6,7,8,9\\

0145$-$257	&26,200	&7.93	&14.59	&78	&1,6,7\\

0205$+$133	&57,400	&7.63	&15.04	&221	&1,2,3\\

0208$-$153	&25,000	&7.9	&15.64	&122	&1,2\\

0219$+$282	&25,000	&7.9	&17.03	&231	&1,19\\

0237$+$115	&70,000	&8.00	&15.96	&272	&1,8,20\\    

0303$-$007	&17,000	&8.0	&16.48	&126	&1,11\\

0324$+$738	&7200	&8.4	&17.05	&40	&1,3,21\\

0347$-$137	&21,300	&8.27	&15.32	&71	&1,4,5,10\\

0354$+$463	&8000	&8.0	&15.58	&33	&1,4,5\\

0357$-$233	&50,000	&7.8	&15.93	&278	&1,4,5,22\\

0458$-$662	&20,000	&7.9	&17.72	&258	&1,23\\

0949$+$451	&14,000	&8.0	&15.77	&77	&1,24\\

1051$+$516	&20,000	&7.9	&17.00	&185	&1,12\\

1133$+$489	&47,500	&7.8	&17.08	&429	&1,18\\

1218$+$497	&35,700	&7.87	&16.39	&254	&1,2\\

1236$-$004	&34,000	&7.9	&17.58	&437	&1,25\\

1247$+$550	&4050	&7.57	&17.79	&25	&26\\

1333$+$005	&8500	&8.0	&17.46	&87	&1,4,5\\

1333$+$487	&14,000	&8.0	&14.10	&35	&1,27\\

1339$+$606	&43,000	&7.68	&16.94	&402	&1,28\\

1412$-$049	&40,000	&7.8	&16.74	&333	&1\\

1419$+$576	&35,000	&7.9	&17.22	&373	&1,12\\

1433$+$538	&22,400	&7.80	&16.12	&151	&1,3,4,5\\

1435$+$370	&25,000	&7.9	&16.63	&192	&1,3\\

1443$+$336	&29,800	&7.83	&16.59	&278	&1,2\\

1458$+$171	&22,000	&7.43	&16.30	&216	&1,2\\

1502$+$349	&20,000	&7.9	&16.62	&156	&1\\

1504$+$546	&25,000	&7.9	&16.20	&158	&1\\

1517$+$502	&31,100	&7.84	&17.80	&413	&1,12,29\\

1558$+$616	&25,000	&7.9	&17.01	&229	&1\\

1603$+$125	&\nodata&\nodata&15.6	&1660	&1,11\\

1619$+$525	&18,000	&7.90	&15.60	&93	&1,2\\

1619$+$414	&20,000	&7.9	&17.14	&198	&1,11\\

1622$+$323	&68,300	&7.56	&16.55	&520	&1,2\\

1631$+$781	&39,900	&7.88	&13.21	&57	&1,4,5,8\\

1646$+$062	&29,900	&7.98	&16.12	&175	&1,2\\

1845$+$683	&37,000	&8.21	&15.30	&116	&1,6,7,9\\

2009$+$622	&25,900	&7.70	&15.26	&134	&1,4,5,30\\

2151$-$015	&8500	&8.0	&14.41	&21	&1,3,4,5\\

2211$+$372	&6300	&8.0	&16.70	&50	&3,33\\

2237$-$365	&7200	&8.0	&17.25	&59	&1,31\\

2317$+$268	&25,000	&7.9	&16.54	&185	&1,32\\

2323$+$256	&6000	&8.0	&17.06	&37	&1,3\\

2349$-$283	&17,300	&7.73	&15.44	&90	&1,10\\
		
\enddata

\tablecomments{A single digit following the decimal place for
log $g$ indicates an assumption of $M=0.60$ $M_{\odot}$.  $V$ magnitudes
are uncontaminated or rederived values based on effective temperature and
models, magnitudes and colors in other filters, or photographic magnitudes
and colors (see \S3.6).  Absolute magnitudes come from the models of
\citet{ber95a,ber95b} as well as the specified references.}

\tablerefs{
(1) This work;
(2) \citealt{lie05};
(3) \citealt{mcc99};
(4) \citealt{far04};
(5) \citealt{far05};
(6) \citealt{fin97};
(7) \citealt{ven97};
(8) \citealt{gre00};
(9) \citealt{nap99};
(10) \citealt{koe01};
(11) \citealt{weg90};
(12) \citealt{ste01};
(13) \citealt{sil02};
(14) \citealt{bra95};
(15) \citealt{gre74};
(16) \citealt{lam00};
(17) \citealt{egg78};
(18) \citealt{wes85};
(19) \citealt{dar96};
(20) \citealt{dre96};
(21) \citealt{gre84};
(22) \citealt{gre79};
(23) \citealt{hut96};
(24) \citealt{jor93};
(25) \citealt{kle04};
(26) \citealt{ber01};
(27) \citealt{dah82};
(28) \citealt{mar97};
(29) \citealt{lie94};
(30) \citealt{ber92};
(31) \citealt{fre00};
(32) \citealt{osw84};
(33) \citealt{kaw04}}

\end{deluxetable}

\clearpage

\begin{deluxetable}{lccccc}
\tablecaption{Parameters of Resolved Secondary \& Tertiary Stars\label{tbl3}}
\tablewidth{0pt}
\tablehead{
\colhead{Primary}	&
\colhead{Companion}	&
\colhead{SpT}		&
\colhead{$I-K$}		&
\colhead{$d_{\rm rd}$ (pc)}	&
\colhead{Ref}}

\startdata

0034$-$211	&LTT 0329B	&dM3.5	&2.24	&51	&1,2,3,4,7,13\\

0116$-$231	&GD 695B	&dM4.5	&2.39	&171	&1,2,14,15\\

0131$-$163	&GD 984B	&dM3.5	&2.24	&110	&1,2,3,4,6,7\\

0145$-$257	&GD 1401B	&dM3.5	&2.29	&79	&1,2,6,16\\

0205$+$133	&PG 0205$+$133B	&dM1	&1.98	&212	&1,2,5,17\\

0208$-$153	&WD 0208$-$153B	&dM2	&2.05	&143	&1,2\\

0219$+$282	&KUV 0219$+$282B&dM5.5	&2.87	&231	&1,2\\

0237$+$115	&PG 0237$+$115B	&dM3	&2.14	&211	&1,5,10,19\\

0324$+$738	&G221-11A	&dM5	&2.65	&40	&1\\
		&G221-11B	&dM6	&3.19	&40	&1\\

0347$-$137	&GD 51B		&dM4.5	&2.40	&52	&1,3,4\\

0357$-$233	&Ton S 392B	&dM3	&2.17	&410	&1,3,4\\

0949$+$451	&HS 0949$+$451B	&dM4.5	&2.53	&66	&1,9\\
		&HS 0949$+$451C	&dM4.5	&2.53	&66	&1,9\\

1133$+$489	&PG 1133$+$489B	&dM5	&2.83	&327	&1,5,18\\

1218$+$497	&PG 1218$+$497B	&dM4	&2.33	&164	&1,2,5,10\\

1236$-$004	&WD 1236$-$004B	&dM4	&2.56	&363	&1,2,20\\

1333$+$487	&GD 325B	&dM5	&2.53	&35	&1,21\\

1412$-$049	&PG 1412$-$049B	&dM0	&1.82	&378	&1,2\\

1419$+$576	&SBS 1419$+$576B&dM2	&2.00	&374	&1,8\\
		&SBS 1419$+$576C&dM2	&2.00	&374	&1,8\\

1435$+$370	&CBS 194B	&dM2.5	&2.09	&192	&1,2\\

1443$+$336	&PG 1443$+$336B	&dM2.5	&2.05	&277	&1,2\\

1502$+$349	&CBS 223B	&dM5	&2.78	&194	&1,2\\

1558$+$616	&HS 1558$+$616B	&dM4.5	&2.43	&136	&1,2,12\\

1619$+$525	&PG 1619$+$525B	&\nodata&\nodata&93	&1,2\\
		&PG 1619$+$525C	&\nodata&\nodata&93	&1,2\\

1619$+$414	&KUV 1619$+$414B&dM5	&2.70	&105	&1,2,22\\

1622$+$323	&PG 1622$+$323B	&dM1	&2.03	&488	&1,2,5\\

1631$+$781	&RE 1631$+$781B	&dM3	&2.18	&85	&1,2,3,4,6,7,11\\
		&RE 1631$+$781C	&dM3	&2.18	&85	&1,2,3,4,6,7,11\\

1646$+$062	&PG 1646$+$062B	&dM3.5	&2.21	&170	&1,5\\

2151$-$015	&LTT 8747B	&dM8	&3.85	&20	&1,3,4\\

\enddata

\tablerefs{
(1) This work;
(2) \citealt{wac03};
(3) \citealt{far04};
(4) \citealt{far05};
(5) \citealt{gre86};
(6) \citealt{gre00};
(7) \citealt{sch96};
(8) \citealt{ste01};
(9) \citealt{jor93};
(10) \citealt{wes85};
(11) \citealt{coo92};
(12) \citealt{mcc99};
(13) \citealt{pro83};
(14) \citealt{egg78};
(15) \citealt{lam00};
(16) \citealt{mue87};
(17) \citealt{wil01};
(18) \citealt{van05};
(19) \citealt{dre96};
(20) \citealt{kle04};
(21) \citealt{gre75};
(22) \citealt{weg90}}

\end{deluxetable}

\clearpage

\begin{deluxetable}{lccccc}
\tablecaption{Unresolved Secondary Parameters\label{tbl4}}
\tablewidth{0pt}
\tablehead{
\colhead{Primary}	&
\colhead{Companion}	&
\colhead{SpT}		&
\colhead{$I-K$}		&
\colhead{$d$ (pc)}	&
\colhead{Ref}}

\startdata

0023$+$388	&G171-B10C	&dM5.5	&2.83	&76	&1,2,6\\

0303$-$007	&KUV 0303$-$007B&dM4	&2.32	&84	&1,2,7\\

0354$+$463	&Rubin 80B	&dM7	&3.47	&41	&1,3,4,5\\

0458$-$662	&WD 0458$-$662B	&dM2.5	&2.12	&171	&1,8\\

1051$+$516	&SBS 1051$+$516B&dM3	&2.20	&183	&1,2\\

1333$+$005	&LP 618-14B	&dM4.5 	&2.52	&147	&1,3,4,5\\

1339$+$606	&RE 1339$+$606B	&dM4	&2.36	&259	&1,9\\

1433$+$538	&GD 337B	&dM5	&2.54	&162	&1,3,4,5,10\\

1458$+$171	&PG 1458$+$171B	&dM5	&2.72	&156	&1,2\\

1504$+$546	&CBS 301B	&dM4	&2.33	&112	&1,2,11\\

1517$+$502	&CBS 311B	&dC	&2.87	&413	&1,2,12\\

1603$+$125	&KUV 1603$+$125B&dK3	&1.38	&1660	&1,2\\

2009$+$622	&GD 543B	&dM4.5	&2.51	&156	&1,3,4,5,13,14\\

2237$-$365	&LHS 3841B	&dM2	&1.89	&282	&1,2\\

2317$+$268	&KUV 2317$+$268B&dM3.5	&2.23	&219	&1,2\\

\enddata

\tablerefs{
(1) This work;
(2) \citealt{mcc99};
(3) \citealt{wac03};
(4) \citealt{far04};
(5) \citealt{far05};
(6) \citealt{rei96};
(7) \citealt{weg87};
(8) \citealt{hut96};
(9) \citealt{fle96};
(10) \citealt{gre75};
(11) \citealt{ste01};
(12) \citealt{lie94};
(13) \citealt{gre84};
(14) \citealt{mor05}}

\end{deluxetable}

\clearpage

\begin{deluxetable}{lccc}
\tablecaption{Projected Separations for All Double Stars\label{tbl5}}
\tablewidth{0pt}
\tablehead{
\colhead{Binary}	&
\colhead{$a_{sky} (''$)}&
\colhead{$a$\tablenotemark{\dag} (AU)}	&
\colhead{Type}}

\startdata

0023$+$388AC	&$<0.025$	&$<1.6$		&WD+RD\\

0034$-$211AB	&0.328		&21		&WD+RD\\

0116$-$231AB	&1.105		&180		&WD+RD\\	
		
0131$-$163AB	&0.189		&21		&WD+RD\\
		
0145$-$257AB	&2.295		&180		&WD+RD\\
		
0205$+$133AB	&1.257		&280		&WD+RD\\
		
0208$-$153AB	&2.647		&320		&WD+RD\\
		
0219$+$282AB	&0.117		&27		&WD+RD\\
		
0237$+$115AB	&0.124		&34		&WD+RD\\
		
0303$-$007AB	&$<0.025$	&$<3.2$		&WD+RD\\
		
0324$+$738BC	&0.297		&12		&RD+RD\\
		
0347$-$137AB	&1.052		&75		&WD+RD\\
		
0354$+$463AB	&$<0.025$	&$<0.8$		&WD+RD\\
		
0357$-$233AB	&1.190		&330		&WD+RD\\
		
0458$-$662AB	&$<0.025$	&$<8.3$		&WD+RD\\
		
0949$+$451AB	&2.892		&220		&WD+RD\\
0949$+$451BC	&0.009		&0.7		&RD+RD\\
		
1051$+$516AB	&$<0.025$	&$<4.6$		&WD+RD\\
				
1133$+$489AB	&0.094		&28		&WD+RD\\
		
1218$+$497AB	&0.302		&77		&WD+RD\\
		
1236$-$004AB	&0.658		&29		&WD+RD\\
		
1333$+$005AB	&$<0.025$	&$<2.2$		&WD+RD\\
		
1333$+$487AB	&2.947		&100		&WD+RD\\
		
1339$+$606AB	&$<0.025$	&$<10$		&WD+RD\\
		
1412$-$049AB	&3.508		&1200		&WD+RD\\
		
1419$+$576AB	&0.658		&250		&WD+RD\\
1419$+$576BC	&0.008		&3.0		&RD+RD\\
		
1433$+$538AB	&$<0.025$	&$<3.8$		&WD+RD\\
		
1435$+$370AB	&1.251		&240		&WD+RD\\
		
1443$+$336AB	&0.679		&190		&WD+RD\\
		
1458$+$171AB	&$<0.025$	&$<5.4$		&WD+RD\\
		
1502$+$349AB	&1.913		&300		&WD+RD\\
		
1504$+$546AB	&$<0.025$	&$<4.0$		&WD+RD\\
		
1517$+$502AB	&$<0.025$	&$<10$		&WD+RD\\
		
1558$+$616AB	&0.715		&160		&WD+RD\\
		
1603$+$125AB	&$<0.025$	&$<42$		&SD+RD\\
			
1619$+$525AB	&2.596		&240		&WD+RD\\
1619$+$525AC	&0.466		&43		&WD+RD\\
		
1619$+$414AB	&0.231		&46		&WD+RD\\
		
1622$+$323AB	&0.094		&49		&WD+RD\\
		
1631$+$781AB	&0.302		&17		&WD+RD\\
1631$+$781BC	&0.007		&0.4		&RD+RD\\
		
1646$+$062AB	&0.163		&29		&WD+RD\\
		
2009$+$622AB	&$<0.025$	&$<3.4$		&WD+RD\\
		
2151$-$015AB	&1.082		&23		&WD+RD\\
		
2237$-$365AB	&$<0.025$	&$<1.5$		&WD+RD\\
		
2317$+$268AB	&$<0.025$	&$<4.6$		&WD+RD\\
	
\enddata

\tablenotetext{\dag}{Values are the current projected separations, not
the true length of the semimajor axes, and are based upon the photometric
distance to the white dwarf (\S3.6).}

\end{deluxetable}


\begin{thebibliography}{}

\bibitem[Bergeron et al.(2001)]{ber01} Bergeron, P., Leggett, S., \& Ruiz,
	M. 2001, \apjs, 133, 413
	
\bibitem[Bergeron et al.(1992)]{ber92} Bergeron, P., Saffer, R., \& Liebert,
	J. 1992, \apj, 394, 228
	
\bibitem[Bergeron et al.(1995a)]{ber95a} Bergeron, P., Saumon, D., \&
	Wesemael, F. 1995a, \apj, 443, 764

\bibitem[Bergeron et al.(1995b)]{ber95b} Bergeron, P., Wesemael, F.,
	\& Beauchamp, A. 1995b, \pasp, 107, 1047
	
\bibitem[Bleach et al.(2002)]{ble02} Bleach, J., Wood, J., Smalley, B.,
	\& Catal\'an, M. 2002, \mnras, 336, 611
	
\bibitem[Bond(1985)]{bon85} Bond, H. 1985, Proceedings of the 7th N.
	American Workshop on Cataclysmic Variables and Low-Mass X-Ray Binaries,
	(Dordrecht: D. Reidel), 15

\bibitem[Bond \& Livio(1990)]{bon90} Bond, H., \& Livio, M. 1990, \apj,
	355, 568

\bibitem[Bragaglia et al.(1995)]{bra95} Bragaglia, A., Renzini, 
	A., \& Bergeron, P. 1995, \apj, 443, 735
	
\bibitem[Catal\'an et al.(1995)]{cat95} Catal\'an, M., Sarna, M.,
	Jomaron, C., \& Smith, R. 1995, \mnras, 275, 153
	
\bibitem[Cooke et al.(1992)]{coo92} Cooke, B., et al. 1992, \nat,
	335, 61

\bibitem[Cutri et al.(2003)]{cut03} Cutri, R., et al. 2003, 2MASS
	All Sky Catalog of Point Sources (IPAC/CIT)

\bibitem[Dahn et al.(1982)]{dah82} Dahn, C., et al. 1982, \aj, 87, 419

\bibitem[Dahn et al.(2002)]{dah02} Dahn, C., et al. 2002, \aj, 124, 1170

\bibitem[Darling \& Wegner(1996)]{dar96} Darling, G., \& Wegner, G.,
	1996, \aj, 111, 865
	
\bibitem[de Kool \& Ritter(1993)]{dek93} de Kool, M., \& Ritter, H.,
	1993, \aap, 267, 397
	
\bibitem[Dreizler \& Werner(1996)]{dre96} Dreizler, S., \& Werner
	K. 1996, \aap, 314, 217
	    
\bibitem[Eggen \& Bessell(1978)]{egg78} Eggen, O., \& Bessell, M.
	1978, \apj, 226, 411

\bibitem[Farihi(2004)]{far04} Farihi, J. 2004, Ph.D. Thesis, UCLA

\bibitem[Farihi et al.(2005)]{far05} Farihi, J., Becklin, E., \&
	Zuckerman, B. 2005, \apjs, 161, 394

\bibitem[Fleming et al.(1996)]{fle96} Fleming, T., Snowden, S., Pfefferman,
	E., Briel, U., \& Greiner, J. 1996, \aap, 316, 147
 
\bibitem[Finley et al.(1997)]{fin97} Finley, D., Koester, D., \& Basri, G.
	1997, \apj, 488, 375
 	
\bibitem[Ford et al.(1998)]{for98} Ford, H., et al. 1998, SPIE,
	3356, 234

\bibitem[Friedrich et al.(2000)]{fre00} Friedrich, S., Koester, D., 
	Christlieb, N., Reimers, D., \& Wisotzki, L., 2000,\aap
	363, 1040

\bibitem[Fuhrmeister \& Schmitt(2003)]{fuh03} Fuhrmeister, B., \&
	Schmitt, J. 2003, \aap, 403, 247
	
\bibitem[Gizis(1997)]{giz97} Gizis, J. 1997, \aj, 113, 806

\bibitem[Green et al.(2000)]{gre00} Green, P., Ali, B., \&
 	Napiwotzki, R. 2000, \apj, 540, 992
 	
\bibitem[Greenstein(1974)]{gre74} Greenstein, J. 1974, \apj, 189,
	L131
	
\bibitem[Greenstein(1975)]{gre75} Greenstein, J. 1975, \apj, 196,
	L117.

\bibitem[Greenstein(1979)]{gre79} Greenstein, J. 1979, \apj, 227,
	244

\bibitem[Greenstein(1984)]{gre84} Greenstein, J. 1984, \apj, 276,
	602

\bibitem[Greenstein(1986)]{gre86} Greenstein, J. 1986, \aj, 92, 867

\bibitem[Hambly et al.(2001)]{ham01} Hambly, N., Irwin, M., \&
	MacGillivray, H. 2001, \mnras, 326, 1295
 	
\bibitem[Holberg et al.(2003)]{hol03} Holberg, J., Barstow, M., \&
	Burleigh, M. 2003, \apjs, 147, 145

\bibitem[Hutchings et al.(1996)]{hut96} Hutchings, J., Crampton, D.,
	Cowley, A., Schmidtke, P., McGrath, T., \& Chu, Y. 1995,
	\pasp, 107, 931

\bibitem[Jeans(1924)]{jea24} Jeans, J. 1924, \mnras, 85, 2

\bibitem[Jordan \& Heber(1993)]{jor93} Jordan, S., \& Heber, U.
	1993, Proceedings of the $8^{th}$ European Workshop on
	White Dwarfs, ed. M. Barstow (Dordrecht: Kluwer), 403, 47

\bibitem[Kawka et al.(2004)]{kaw04} Kawka, A., Vennes, S., \&
	Thorstensen, J. 2004, \aj, 127, 1702
	
\bibitem[Kirkpatrick \& McCarthy(1994)]{kir94} Kirkpatrick, J., \& McCarthy,
	D. 1994, \aj, 107, 333
	
\bibitem[Kleinman et al.(2004)]{kle04} Kleinman, S. et al. 2004,
	\apj, 607, 426

\bibitem[Koester et al.(2001)]{koe01} Koester, D., et al. 2001,
	\aap, 378, 556

\bibitem[Lamontagne et al.(2000)]{lam00} Lamontagne, R., Demers, S.,
	Wesemael, F., Fontaine, G., \& Irwin, M. 2000, \aj, 119, 241

\bibitem[Liebert et al.(2005)]{lie05} Liebert, J., Bergeron, P.,
 	\& Holberg, J. 2004, \apjs, 156, 47
 	
\bibitem[Liebert et al.(1994)]{lie94} Liebert, J., et al. 1994, \apj.
	421, 733	

\bibitem[Livio \& Soker(1984)]{liv84} Livio, M., \& Soker, N. 1984,
	\mnras, 208, 783
	
\bibitem[Livio \& Soker(1988)]{liv88} Livio, M., \& Soker, N. 1988,
	\apj, 329, 764
	
\bibitem[Livio(1996)]{liv96} Livio, M. 1996, ASP Conference Series 90:
	The Origins, Evolution, and Destinies of Binary Stars in Clusters,
	ed. E. Milone, J. Mermilliod, (San Francisco: ASP), 291
        
\bibitem[Marsh et al.(1995)]{mar95} Marsh, T., Dhillon, V., \&
 	Duck, S. 1995, \mnras, 275, 828
 	
 \bibitem[Marsh et al.(1997)]{mar97} Marsh, M., et.al. 1997, \mnras,
	286,369

\bibitem[Maxted et al.(1998)]{max98} Maxted, P., Marsh, T., Moran, C.,
	Dhillon, V., \& Hilditch, R. 1998, \mnras, 300, 1225
	
\bibitem[Maxted et al.(2000)]{max00} Maxted, P., Moran, C., Marsh, T., \&
 	Gatti, A. 2000, \mnras, 311, 877
 	
\bibitem[McCook \& Sion(1999)]{mcc99} McCook, G., \& Sion, E. 1999,
	\apjs, 121, 1

\bibitem[Monet et al.(1998)]{mon98} Monet, D., et al. 1998, The
	USNO-A2.0 Catalogue (U.S. Naval Observatory Flagstaff Station
	and Universities Space Research Association)
	
\bibitem[Monet et al.(2003)]{mon03} Monet, D., et al. 2003, \aj, 125,
	984

\bibitem[Morales-Rueda et al.(2005)]{mor05} Morales-Rueda, L., Marsh, T.,
	Maxted, P., Nelemans, G., Karl, C., Napiwotzki, R., \& Moran, C.
	2005, \mnras, 359, 648
		
\bibitem[Mueller \& Bues(1987)]{mue87} Mueller, B., \& Bues, I.
	MitAG, 70, 345
	
\bibitem[Napiwotzki et al.(1999)]{nap99} Napiwotzki, R., Green, P.,
	\& Saffer, R. 1999, \apj, 517, 399

\bibitem[Oswalt et al.(1984)]{osw84} Oswalt, T., Peterson, B., \&
	Foltz, C. 1984, \aj, 89, 421

\bibitem[Paczynski(1976)]{pac76} Paczynski, B. 1976, Proceedings of
	IAU Symposium 73, eds. P. Eggleton, S. Mitton, \& J. Whelan
	(Dordrecht: D. Reidel), 75

\bibitem[Probst(1983)]{pro83} Probst, R. 1983, \apjs, 53, 335
	
\bibitem[Raymond et al.(2003)]{ray03} Raymond, S., et al. 2003, \aj,
	125, 2621

\bibitem[Reid(1996)]{rei96} Reid, I. 1996, \aj, 111, 2000

\bibitem[Reid \& Hawley(2000)]{rei00} Reid, I., \& Hawley, S. 2000,
	in New Light on Dark Stars, (New York: Springer)
 	
\bibitem[Saffer et al.(1993)]{saf93} Saffer, R., Wade, R., Liebert,
	J., Green, R., Sion, E., Bechtold, J., Foss, D., \& Kidder, K.
	1993, \aj, 105, 1945
	
\bibitem[Schmidt et al.(1995)]{sch95} Schmidt, G., Smith, P., \&
	Harvey, D. 1995, \aj, 110, 398
	
\bibitem[Schreiber \& G\"ansicke(2003)]{sch03} Schreiber, M., \& G\"ansicke,
	B. 2003, \aap, 406, 305

\bibitem[Schultz et al.(1996)]{sch96} Schultz, G., Zuckerman,
	B., \& Becklin E. 1996, \apj, 460, 402

\bibitem[Silvestri et al.(2002)]{sil02} Silvestri, N., Oswalt,
 	T., \& Hawley, S. 2002, \aj, 124, 1118

\bibitem[Sion et al.(1995)]{sio95} Sion, E., Holberg, J., Barstow, M.,
	\& Kidder, K. 1995, \pasp, 107, 232

\bibitem[Sirianni et al.(2005)]{sir05} Sirianni, M. et al. 2005,
	PASP, 117, 1049

\bibitem[Stepanian et al.(2001)]{ste01} Stepanian, J., Green, R., Foltz, 
	C., Chaffee, F., Chavushyan, V., Lipovetsky, V., \& Erastova, L.
	2001, \aj, 122, 3361

\bibitem[Valls-Gabaud (1988)]{val88} Valls-Gabaud, E. 1988, \apss, 142, 289

\bibitem[van den Besselaar et al.(2005)]{van05}	van den Besselaar, E.,
	Roelofs, G., Nelemans, G., Augusteijn, T., \& Groot, P. 2005,
	\aap, 434, L13
	
\bibitem[Vennes et al.(1997)]{ven97} Vennes, S., Thejll, P., Galvan, R., \&
	Dupuis, J. 1997, \apj, 480, 714
	
\bibitem[Wachter et al.(2003)]{wac03} Wachter, S., Hoard, D., Hansen,
	K., Wilcox, R., Taylor, H., \& Finkelstein, S. 2003, \apj,
	586, 1356
	
\bibitem[Wegner et al.(1990)]{weg90} Wegner, G., Africano, J., \& Boodrich,
	B. 1990, \aj, 99, 1907
	 
\bibitem[Wegner et al.(1987)]{weg87} Wegner, G., McMahan, R., Boley,
	\& Forrest I. 1987, \aj, 94, 1271
	
\bibitem[Wesemael et al.(1985)]{wes85} Wesemael, F., Green, R., \&
	Liebert, J. 1985, \apjs, 58, 379
    
\bibitem[Williams et al.(2001)]{wil01} Williams, T., McGraw, J., \&
	Grashuis, R. 2001, \pasp, 113, 490 
	
\bibitem[Yungelson et al.(1993)]{yun93} Yungelson, L., Tutukov, A.,
	\& Livio, M. 1993, \apj, 418, 794
    
\bibitem[Zuckerman \& Becklin(1987)]{zuc87} Zuckerman, B., \& Becklin,
	E. 1987, \apj, 319, 99
	
\bibitem[Zuckerman \& Becklin(1987b)]{zuc87b} Zuckerman, B., \& Becklin, E.
	1987b, \nat, 330.  138

\bibitem[Zuckerman \& Becklin(1992)]{zuc92} Zuckerman, B., \& Becklin,
	E. 1992, \apj, 386, 260

\end{thebibliography}
\end{document}